%% file: main.tex
\documentclass[9pt,twocolumn,twoside]{osajnl}

\journal{jocn}

\setboolean{shortarticle}{false}

\usepackage{amsmath,amssymb,amsfonts}

\usepackage{algpseudocode}
\usepackage{algorithmicx}
\usepackage{algorithm}

\usepackage{tabularx}
\usepackage{amsmath}

\usepackage{booktabs}
\usepackage[utf8]{inputenc}
\usepackage[T1]{fontenc}
\usepackage[english]{babel}
\usepackage{amsthm}
\theoremstyle{definition}
 
\newboolean{withappendix}
\setboolean{withappendix}{True}

\let\emptyset\varnothing

\usepackage{graphicx}
\usepackage{textcomp}
\usepackage[utf8]{inputenc}
\usepackage{multirow}
\usepackage{float} 
\usepackage{xcolor}
\def\BibTeX{{\rm B\kern-.05em{\sc i\kern-.025em b}\kern-.08em
    T\kern-.1667em\lower.7ex\hbox{E}\kern-.125emX}}

\usepackage{mathtools}

\usepackage{wrapfig}
\usepackage{graphicx}
\usepackage{subcaption}
\usepackage{float}

\usepackage{epstopdf}
\usepackage[xdvi]{epsfig} 

\usepackage{float} 

\usepackage{color,soul}

\usepackage{adjustbox}
\usepackage{multirow}
\usepackage{tablefootnote}
\usepackage{algorithm}

\title{A benchmarking framework for PON-based fronthaul network design}

\author[1,*]{Egemen Erbayat}
\author[2]{Gustavo B. Figueiredo}
\author[3]{Shih-Chun Lin}
\author[4]{Motoharu Matsuura}
\author[5]{Hiroshi Hasegawa}
\author[1]{Suresh Subramaniam}

\affil[1]{The George Washington University, USA, $^{2}$Federal University of Bahia, Brazil, $^{3}$North Carolina State University, USA, $^{4}$University of Electro-Communications, Japan, $^{5}$Nagoya University, Japan}

\affil[*]{C\textbf{}orresponding author: erbayat@gwu.edu}

\begin{abstract}
As mobile networks transition toward 5G and 6G RAN architectures, Passive Optical Networks (PONs) offer a critical solution for cost-effective fronthaul transport. However, the lack of standardized evaluation models in current literature makes an objective comparison of diverse optimization strategies difficult. This paper addresses this gap by proposing a unified benchmarking framework that standardizes cost catalogs and deployment scenarios. We formulate the network design problem using Integer Linear Programming (ILP) to establish optimality bounds and evaluate three scalable heuristic strategies: a Genetic Algorithm, K-Means Clustering (KMC+), and a graph-based Randomized Successive Splitter Assignment (RSSA+) algorithm. Simulation results show that a time-limited ILP remains a strong reference point, even when optimality is not reached. Despite being rarely used in prior fronthaul planning studies, it consistently yields solutions superior to those produced by standard heuristic methods. Among scalable approaches, RSSA+ reliably attains near-ILP performance while ensuring feasibility across all evaluated scenarios, which underscores the importance of advanced, constraint-aware algorithmic designs over simpler heuristics.
\end{abstract}

\begin{document}
\maketitle

\input{sections/introduction}

\input{sections/background}

\input{sections/framework}

\input{sections/methods}

\input{sections/evaluation}

\input{sections/results}

\input{sections/conclusion}

\section*{Acknowledgment}
This work was supported in part by NSF grant CNS-2210343.

\bibliography{references}

\ifthenelse{\boolean{withappendix}}
{\appendix
\section*{Appendix}
\input{sections/appendix}}{}

\end{document}

%% file: sections/introduction
\section{Introduction}

The unprecedented demand for ultra-high-speed, low-latency, and reliable connectivity is reshaping mobile network architecture as the industry transitions from 5G to the vision of 6G \cite{6GVision, 10546919}. Driven by applications such as immersive extended reality (XR), autonomous mobility, massive Internet of Things (IoT), and the tactile Internet, future wireless systems must support orders-of-magnitude growth in data volume, device density, and service diversity \cite{ITU-R-2370}. To meet these performance goals, Radio Access Networks (RANs) are becoming increasingly disaggregated, with baseband functions centralized and cell deployments densified. While Centralized or Cloud-RAN (C-RAN) architectures provide enhanced spectral efficiency and lower operational cost, they place heavy demands on the fronthaul segment linking remote radio heads (RRHs) with centralized baseband units (BBUs) \cite{Checko2015,cpri_specification_70,ecpri_specification_20}. As fronthaul links are required to simultaneously provide multi-gigabit throughput, sub-millisecond latency, and synchronization precision, transport acts as a critical bottleneck for scalable 5G and 6G deployments \cite{ORAN-WG4, 3GPP-38801,sup6620205g}.

Passive Optical Networks (PONs) provide a high-capacity, low-latency, and cost-effective solution for delivering fiber-based transport from the central office to numerous cell towers \cite{horvath2020passive}. Since PONs facilitate shared fiber infrastructure and avoid the need for a dedicated fiber per tower, they are uniquely suitable for scenarios where multiple cell sites need to be served economically \cite{maes2024efficient}. Modern standards such as 10-Gigabit-capable Symmetric PON (XGS-PON), Next-Generation PON 2 (NG-PON2), and 50-Gigabit-capable PON (50G-PON) offer bandwidths up to 10–50 Gbps per wavelength and support time and/or wavelength division multiplexing \cite{ITUTG989, ITUTG9804, ITU_T_G9807, Nakayama2019}. Consequently, they are well-suited to carry fronthaul traffic for low-layer functional splits such as 3rd Generation Partnership Project (3GPP) Option 7.2 \cite{11021422}.

A wide range of research studies have examined the utilization of PONs for fronthaul within 4G, 5G, and anticipated 6G networks. While some works focus on physical-layer feasibility by showing analog and digital transmission of radio signals across PON topologies \cite{11006330, 11172732}, others introduce optimization models for planning and dimensioning cost-effective and low-latency PON deployments \cite{Chen2016JOCN, ranaweera20175g, ranaweera2019optical, musumeci2016optimal, dias2022evolutionary,akhtar2023fronthaul, lisi2017cost, masoudi2020cost, carapellese2015bbu, kokangul2011optimization, fayad2022design, fayad20235g, erbayat2024fronthaul, erbayat2024design, erbayat2025multi, erbayat2025toward,erbayat2025latency, erbayat2025design, ciceri2024resource}. These studies use a variety of functional splits and network models to provide a broad perspective. This diversity enriches the field, but it creates context-specific results which are not always suited for direct comparison.

The evolution toward open and disaggregated architectures, such as those promoted by the Open RAN (O-RAN) Alliance \cite{ORAN-WG4}, further intensifies the need for objective assessment of transport solutions. In the remainder of this paper, we adopt the O-RAN terminology of Distributed Unit (DU) and Radio Unit (RU) in place of the traditional BBU and RRH designations, since the two pairs refer to the same logical network functions.\footnote{We retain the BBU and RRH terms only in direct references to prior D-RAN and C-RAN literature.} This capability is crucial to ensure interoperability, scalability, and cost-effectiveness. However, we recognize that specific cost parameters and technical constraints are inherently site-dependent and fluctuate based on the chosen technology and geographical location. This variability creates a significant challenge for comparative research, as differences in reported results often stem from mismatched input assumptions rather than algorithmic performance. Consequently, a key limitation within the current literature is the absence of standardized benchmarking scenarios and unified evaluation frameworks; without these, performance claims remain context-dependent and cannot be fairly compared.

In this paper, we address this gap by proposing the first standardized benchmarking framework for PON-based fronthaul design and enable reproducible comparisons that was previously impossible. Rather than relying on isolated case studies, our framework utilizes a standardized cost catalog derived from broad industry averages to mitigate the impact of parameter variability. By defining fixed functional split scenarios, normalized techno-economic models, and standard deployment templates, we establish a controlled and reproducible environment for evaluation. We further validate this framework by quantitatively benchmarking an exact Integer Linear Programming (ILP) formulation against three scalable heuristic strategies. These include a Genetic Algorithm (GA), an improved K-means clustering method (KMC+), and a graph-based Improved Randomized Successive Splitter Assignment (RSSA+). The latter two build on \cite{erbayat2025toward} and are extended in this study to address practical limitations, with KMC+ introducing a principled mechanism for determining the number of clusters and RSSA+ incorporating cost heterogeneity into the assignment process. Taken together, these methods are evaluated within a unified benchmarking framework. Through this benchmarking, we enable transparent comparisons of PON fronthaul architectures for 5G and beyond and ensure that performance differences remain interpretable across diverse optimization methods.

Beyond comparative evaluation, our results yield several concrete design insights. First, a time-limited ILP, an exact formulation solved under a fixed runtime budget without requiring optimality, serves as a strong and informative baseline. Its suboptimal solutions consistently outperform heuristics and provide meaningful lower bounds. Second, heuristic performance depends strongly on explicit constraint handling and cost awareness. Simple methods such as GA degrade under strict physical constraints, while structured approaches such as RSSA+ and KMC+  closely track ILP performance at scale, with RSSA+ showing greater flexibility under heterogeneity due to its constructive nature. Third, the results expose clear trade-offs between DU availability, site-related costs, and infrastructure sharing that fixed clustering methods cannot capture. Together, these findings reinforce the need for constraint-aware and cost-sensitive optimization and demonstrate the value of standardized benchmarking for fronthaul design in future mobile networks.

%% file: sections/background
\section{Background}
\label{s:related_work}

\subsection{Modern RAN Architectures and Fronthaul Requirements}
Fronthaul networks play a critical role in modern RAN deployments by enabling centralized processing, improved spectral efficiency, and more flexible resource allocation across radio sites. High-capacity, low-latency fronthaul links allow baseband resources to be pooled, reducing hardware duplication, lowering energy consumption, and simplifying network management. These benefits have been a key enabler for the decoupling of hardware and software that underpins contemporary RAN evolution \cite{wypior2022open}.

The traditional Distributed RAN (D-RAN) architecture, characterized by co-located BBUs and RUs, offers low latency but suffers from high operational costs and inefficient resource utilization due to its rigid, distributed nature \cite{habibi2019comprehensive}. C-RAN was introduced to address these inefficiencies by aggregating BBUs into centralized pools to improve spectral efficiency and reduce energy consumption, though this creates stringent low-latency requirements on the fronthaul transport network. Variations such as Heterogeneous Cloud RAN (HC-RAN), Fog RAN (F-RAN), and Virtualized RAN (v-RAN) have since emerged to balance the trade-offs between centralized processing and edge-computing requirements, and these architectures deliver improvements in coverage, latency, and flexibility through network function virtualization \cite{10546919, tinini2020energy}.

By building upon the principles of virtualization, O-RAN represents a transformative shift through standardized open interfaces that effectively break the proprietary vendor lock-in inherent in previous architectures \cite{polese2023understanding}. Furthermore, O-RAN is particularly promising because it fosters a multi-vendor ecosystem where operators can mix and match components from different suppliers, and this capability drives competition and innovation \cite{liyanage2023open}. Additionally, the introduction of the RAN Intelligent Controller (RIC) within the O-RAN architecture enables the integration of Artificial Intelligence (AI) and Machine Learning (ML) for near-real-time network optimization and intelligent resource management, significantly enhancing network agility and reducing long-term capital and operational expenditures \cite{bonati2021intelligence}.

Another important source of architectural flexibility in modern RANs is the set of functional split options defined by 3GPP and adopted by the O-RAN Alliance, which determine how baseband processing is divided between the DU and the RU \cite{3GPP-38801,ORAN-WG4}. These splits span a wide spectrum and range from highly centralized architectures to more distributed designs. Low-layer splits such as Option~8 and Option~7.1 keep most physical-layer processing at the DU and enable tight coordination and centralized control. This design choice comes at the cost of extremely demanding fronthaul requirements. Data rates can reach hundreds of Gbps per RU, and one-way latency budgets are typically constrained to 100--250~$\mu$s \cite{sup6620205g}. These requirements significantly constrain transport design and often necessitate dedicated high-capacity links.

To alleviate these constraints, current deployments increasingly favor Option 7.2x. This split shifts a limited set of physical-layer functions to the RU. As a result, fronthaul bandwidth requirements are reduced to approximately 2-6 Gbps per RU while real-time scheduling and coordination remain centralized at the DU. This combination makes Option 7.2x well suited for optical fronthaul solutions based on PON technologies. In contrast, higher-layer splits such as Options 1-5 move additional MAC and upper-layer functions toward the RU. These splits prioritize transport efficiency by lowering throughput requirements to around 1 Gbps and relaxing latency constraints to the millisecond range of approximately 1.5-10~ms. This approach reduces the degree of centralization and limits advanced coordination capabilities \cite{larsen2018survey, dao2024review}.

\subsection{Fronthaul Transport Technologies and PON Architectures}

A diverse range of technologies exists to address the stringent throughput and latency requirements of 5G and 6G fronthaul \cite{10546919}. Point-to-Point (P2P) fiber remains the gold standard in terms of performance, as it offers massive throughput, yet it suffers from high deployment costs and low flexibility.
 Wireless alternatives, such as microwave and millimeter-wave (mmWave) links, offer easier deployment but are constrained by lower capacities and shorter transmission distances. Free Space Optics (FSO) presents a high-bandwidth wireless alternative with very low latency, though it requires strict line-of-sight and is highly sensitive to weather conditions. Consequently, optical fiber technologies are widely regarded as the most suitable transport medium to meet fronthaul requirements due to their low attenuation, high bandwidth, and scalability \cite{liu2022enabling}. However, deploying dedicated P2P fiber links to every RU is capital-intensive and logistically challenging. PONs offer a cost-effective alternative by using power splitters and a shared optical distribution network (ODN) to connect multiple optical network units (ONUs) at cell sites to a centralized optical line terminal (OLT) located at a central office \cite{horvath2020passive,maes2024efficient}. This architecture supports point-to-multipoint communication, which significantly reduces the number of fiber strands required and allows infrastructure sharing.

Modern PON standards have been progressively enhanced to support higher speeds and service multiplexing. XGS-PON provides symmetric 10~Gbps transmission \cite{ITUTG989}, 
while NG-PON2 introduces multiple 10~Gbps wavelengths with tunable transceivers, enabling dynamic capacity allocation and improved fault tolerance \cite{ITUTG989}. More recently, 25G-PON and 50G-PON technologies, \cite{ITUTG9804}, have been standardized to support emerging fronthaul traffic loads. To optimize performance further, various multiplexing architectures are employed. Time Division Multiplexing PON (TDM-PON) is widely deployed for its simple architecture and cost-effectiveness, though it is limited by distance and bandwidth sharing. Wavelength Division Multiplexing PON (WDM-PON) assigns dedicated wavelengths to ONUs, offering low latency and security at the cost of tunable components, while Time and Wavelength Division Multiplexing PON (TWDM-PON) combines these approaches to achieve capacities exceeding 40~Gbps with flexible bandwidth allocation. Furthermore, advanced schemes such as Orthogonal Frequency Division Multiplexing PON (OFDM-PON) and Non-Orthogonal Multiple Access PON (NOMA-PON) employ complex signal processing to maximize spectral efficiency \cite{feng2023key}.

\subsection{Related Work}

Recent literature on optical fronthaul and C-RAN design exhibits a wide range of system models, spanning fully centralized architectures, partially centralized deployments with functional splitting, and access-focused PON planning. A dominant objective across these studies is the minimization of Total Cost of Ownership (TCO), which combines Capital Expenditures (CapEx) for infrastructure and equipment with Operational Expenditures (OpEx) related to energy, leasing, and maintenance. However, the underlying assumptions shaping TCO formulations differ substantially, which leads to results that are not directly comparable.

Several works focus on the techno-economic feasibility of migrating from distributed RAN to centralized or virtualized C-RAN. \cite{lisi2017cost} and \cite{masoudi2020cost} formulate ILP or Mixed ILP (MILP) models that jointly optimize DU pool placement and fronthaul design under strict latency and capacity constraints. These studies explicitly distinguish between greenfield and brownfield fiber deployments and assume TWDM-PON fronthaul architectures with fixed splitter hierarchies and Common Public Radio Interface (CPRI) based traffic models. Their conclusions regarding the cost effectiveness of full versus partial centralization are therefore tightly coupled to assumed functional split options, fixed per-RU bandwidth demands, and specific cost parameters for fiber rollout and energy consumption.

Other studies emphasize hierarchical aggregation and transport-layer constraints rather than migration economics. \cite{musumeci2016optimal} investigates BBU hotel placement over WDM aggregation networks, focusing on wavelength routing, consolidation gains, and stringent round-trip latency budgets. Their formulation abstracts access-layer splitter placement and does not consider shared PON architectures, which limits comparability with TWDM-PON-based C-RAN studies that explicitly model optical splitting loss and feeder sharing.

Comparative analyses of fronthaul transport technologies introduce further heterogeneity. \cite{ranaweera20175g, ranaweera2019optical} compare CPRI, packetized fronthaul, and analog radio-over-fiber under 5G capacity and latency requirements. While these works develop optimization frameworks to evaluate deployment cost, they assume fixed functional splits and primarily evaluate technology suitability rather than algorithmic performance. As a result, their conclusions depend strongly on assumed traffic models, antenna configurations, and per-cell bandwidth scaling.

More recent work extends fronthaul optimization to dense 5G and early 6G scenarios. \cite{fayad2022design, fayad20235g} analyze TWDM-PON and hybrid PON–FSO architectures under sparse and dense deployments and incorporate energy consumption and link availability constraints.
 \cite{akhtar2023fronthaul} proposes a highly detailed model for option-7 functional splits over WDM-PON, explicitly integrating G/G/1 queuing models to capture upstream delay. While these formulations improve realism, they rely on specific traffic assumptions and queuing parameters, making cross-paper performance comparisons difficult.

A separate body of work concentrates on access network planning independent of RAN virtualization. \cite{dias2022evolutionary} and  \cite{kokangul2011optimization} employ evolutionary and heuristic algorithms for splitter placement and customer assignment using geographic data and optical power budgets. These approaches typically abstract away DU placement, functional splits, and latency constraints, which limits their applicability to end-to-end fronthaul benchmarking.

More recently, Power-over-Fiber (PWoF) technology has emerged as a promising solution to eliminate local power dependencies by transmitting both data and optical power over fiber \cite{de2024radio}. It brings an additional design dimension. To address this, recent studies formulated the "Power-Enabled Optical Fronthaul Design" (POFD) problem and proposed optimization frameworks for splitter placement and routing that account for power delivery constraints, survivability, and latency in both single and multi-DU architectures \cite{erbayat2024fronthaul, erbayat2024design, erbayat2025multi, erbayat2025toward, erbayat2025latency, erbayat2025design}. These studies introduce new constraints and objectives that further differentiate their evaluation settings from conventional PON-based fronthaul models.

Across this literature, network design is consistently framed as a constrained multi-objective optimization problem. However, assumptions regarding functional splits, traffic models, splitter hierarchies, cost parameters, and evaluation metrics vary widely. Algorithms are assessed using heterogeneous datasets, ranging from synthetic grids to real geographic layouts, and performance is reported using incompatible metrics such as TCO crossover time, wavelength usage, consolidation ratios, or heuristic optimality gaps. Consequently, the reported effectiveness of a given algorithm is often inseparable from its modeling assumptions. This diversity highlights the need for a standardized benchmarking framework that fixes architectural assumptions and evaluation criteria, enabling reproducible and fair comparison of fronthaul optimization algorithms.

Lastly, standardized benchmarking methodologies exist in adjacent networking domains, but none addresses PON-based fronthaul design. In optical access planning, prior work fixes datasets and a common cost model to compare splitter placement and customer assignment algorithms \cite{dias2022evolutionary, kokangul2011optimization}, yet abstracts away DU placement, functional splits, and the sub-millisecond fronthaul latency budget. In optical core and metro, recent JOCN efforts standardize transmission parameters, metrics, and baselines for resource allocation in multicore fiber elastic optical networks \cite{pintorios2024benchmarking} and provide the open-source Optical Networking Gym toolkit for QoT-aware dynamic resource assignment \cite{natalino2024gym}, but both target spectrum and wavelength assignment in transport networks rather than the splitter hierarchy, trenching and fiber cost catalog, power budget, and DU/RU placement that drive fronthaul CapEx and OpEx. In the RAN and mobile core, 5GC-Bench \cite{panitsas20255gcbench} and OpenRAN Gym \cite{bonati2022openrangym} benchmark 5G core VNFs and O-RAN xApps respectively at the software and interface layers, while optimization-oriented studies of BBU or DU pool placement \cite{musumeci2016optimal, carapellese2015bbu} each use their own cost catalog and transport abstraction. None of these efforts jointly fixes the PON cost catalog, the reach/latency/optical-power constraints, the DU and splitter capacity ladder, and the service-class parameters, which motivates the dedicated benchmarking framework proposed here.

A parallel line of work on large-scale network planning has developed hybrid exact--heuristic and matheuristic approaches that scale beyond monolithic ILP, including iterative LP with slope scaling for capacitated fixed-charge network design \cite{gendron2018matheuristics}, Benders decomposition with a bees algorithm for transmission expansion planning \cite{macrae2019hybrid}, and hybrid iterated local search with embedded binary-programming subproblems for single-source capacitated facility location \cite{dealmeida2024hybrid}. Telecom- and optical-specific variants embed LP/MIP inside a tabu search for WDM traffic grooming \cite{wu2020matheuristic} and combine improved ILP formulations with a two-stage graph-coloring row-generation heuristic for routing, wavelength, and spectrum assignment \cite{zheng2025improved}. These methods broaden the algorithmic landscape but are each validated on their own instances and cost model, which reinforces the need for a standardized benchmarking environment that allows such strategies to be compared against a common ILP reference.

%% file: sections/framework
\section{Benchmarking Framework Design}\label{sec:benc}

\begin{figure}
    \centering
    \includegraphics[width=1\linewidth]{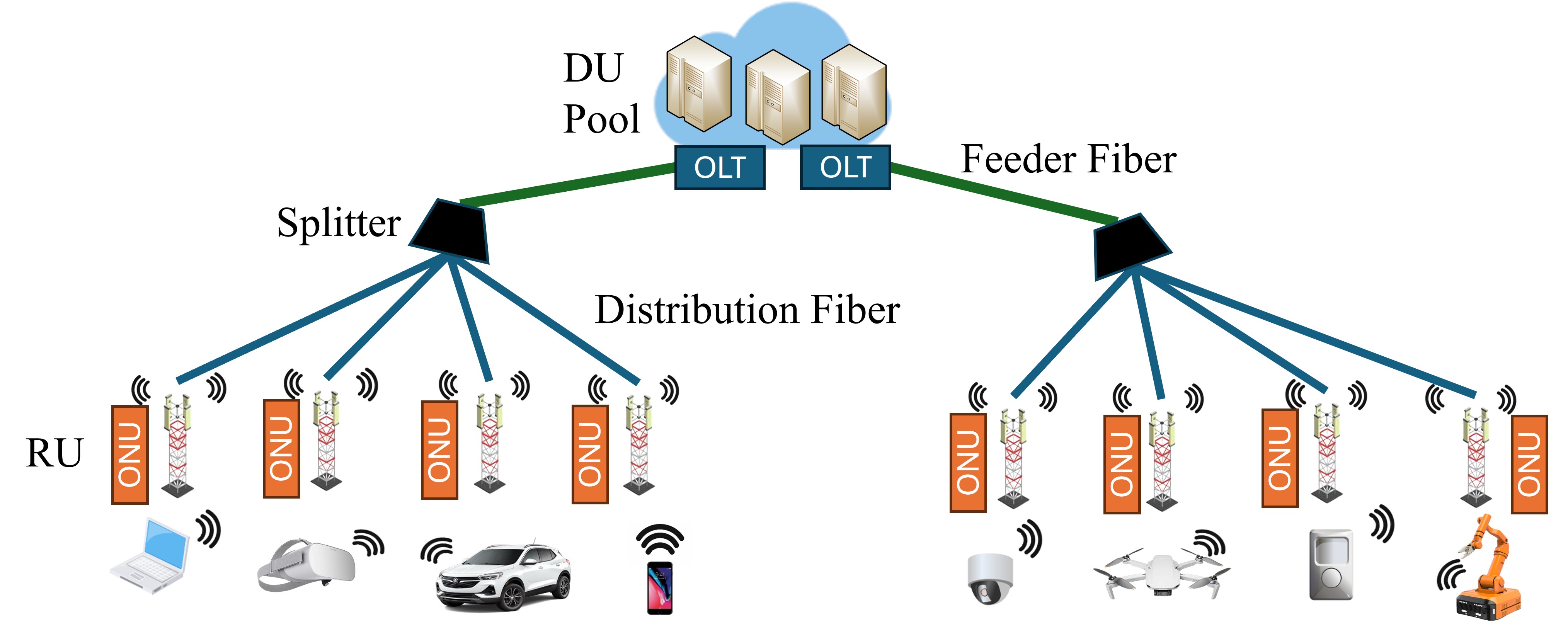}
    \caption{Proposed benchmarking architecture with one DU.}
    \label{fig:system_model}
\end{figure}

\subsection{Common Assumptions}
To ensure that algorithms can be compared on equal footing, we adopt a unified set of assumptions that fix the underlying network environment while allowing each method to differ only in how it selects or optimizes the fronthaul design. We consider a PON-based fronthaul architecture with a predefined set of candidate DU locations and candidate splitter sites, which reflect realistic geographical and regulatory constraints that prevent arbitrary placement \cite{hughes2021planning}. The RU positions are treated as fixed coordinates that are drawn from a representative synthetic layout to ensure that all algorithms operate over the same spatial distribution. To facilitate broad applicability, we generalize the network architecture into three abstract node types—DU, Splitter, and RU—which allows us to simplify the cost model while remaining technology-agnostic. In this approach, the fixed cost assigned to each node encapsulates the aggregate expense of equipment, site acquisition, and installation necessary to deploy that node. This simplification facilitates direct comparison between different algorithms without compromising optimization accuracy, as the dominant cost driver in fronthaul deployment is invariably trenching and fiber rather than specific device variances. Consequently, to address the high cost of civil works, our problem formulation explicitly allows for the deployment of multiple splitters at any splitter location; this design choice reduces total expenditure by maximizing the use of shared trenching paths for multiple connections.

On the technology side, we model a generic fronthaul PON with fixed propagation delay and reach limits. This study focuses on fronthaul deployment rather than physical layer optimization. Therefore, we abstract complex transmission physics into fixed constraints based on industry standards. We restrict the architecture to a single layer of passive splitters between the DU and RUs. This captures the dominant trade-offs between cost and latency without adding unnecessary complexity. The architecture is summarized in Figure \ref{fig:system_model}. We also assume a fixed split option and uniform per-RU bandwidth demand. These assumptions create a unified baseline for benchmarking. They ensure that all algorithms solve the exact same problem instance. However, the underlying model remains adaptable because the input parameters are flexible. Researchers can easily adjust these parameters to reflect the specific latency and bandwidth requirements of other functional splits or architectural constraints. 

While the architectural model establishes the physical structure, a fair comparison also requires a standardized economic environment. Although specific cost parameters and technical constraints are inherently site-dependent, valid benchmarking requires a controlled environment. We therefore synthesize a unified set of assumptions to establish a representative baseline that covers the spectrum of typical deployment scenarios. We adopt a cost catalog based on widely cited industry averages rather than niche case studies. This catalog includes trenching cost per meter, fiber cost per meter, and fixed installation costs for DUs and splitters. This standardization prevents discrepancies caused by heterogeneous cost assumptions in prior studies. It ensures that differences in reported total deployment cost stem from design choices rather than mismatched price inputs. We also apply latency and reach constraints identically across all experiments. We enforce a single one-way fronthaul latency budget and a maximum allowable DU-to-RU distance consistent with ITU PON specifications. For evaluation, every algorithm is tested on the same small, medium, and large-scale scenarios. We assess them using a consistent set of metrics. These include total deployment cost under the shared cost model, number of splitters and DUs used, average RU--splitter distance, fraction of RUs that satisfy latency and reach constraints, and computational runtime. ILP-based methods run with the same solver and time limit to ensure comparable solution quality. Similarly, heuristics operate under consistent computational budgets. These standardized parameters and metrics create a coherent benchmarking framework. Within this framework, performance differences are interpretable and reproducible across diverse fronthaul optimization approaches.

 \subsection{Network Services} \label{sec:network_services}
The optimal design of a PON-based fronthaul network is strictly governed by the service mix it must support. As the industry evolves from 5G toward 6G, where advanced machine learning tasks require stringent latency and efficient computation offloading \cite{liu2023machine, erbayat2024trade}, the network must accommodate three primary service classes \cite{ITURM2412}, each imposing distinct constraints on the architecture:

\begin{itemize} \setlength\itemsep{0em} \setlength\parskip{0em}
\item \textbf{Enhanced Mobile Broadband (eMBB):} Addresses the demand for ultra-high data rates (e.g., immersive XR). This places significant capacity pressure on the transport layer, necessitating high-bandwidth optical standards to ensure the infrastructure can handle heavy data loads.

    \item \textbf{Ultra-Reliable Low-Latency Communications (URLLC):} Driven by mission-critical applications like autonomous mobility, this treats latency as a hard constraint rather than an optimization objective. Processing units are often positioned closer to the network edge to minimize the physical distance signals must travel.

    \item \textbf{Massive Machine-Type Communications (mMTC):} Characterized by high device density but low data rates (e.g., massive IoT), this utilizes point-to-multipoint topologies. Optimization focuses on efficiency to connect a vast number of devices while minimizing physical infrastructure costs.
\end{itemize}

%% file: sections/methods
\section{Optimization Methods Under Study}

We address the challenge of designing a cost-optimal PON-based fronthaul architecture by jointly optimizing the placement of DUs and optical splitters to minimize TCO while satisfying strict latency and capacity requirements. We evaluate four distinct approaches, including an exact ILP formulation, a GA, and two heuristics adapted from prior work. Specifically, we build on previously proposed KMC and RSSA methods in \cite{erbayat2025toward} by introducing improved variants, denoted as KMC+ and RSSA+, which respectively enable adaptive cluster selection and incorporate cost heterogeneity. While NP-hardness makes ILP intractable for large-scale instances \cite{erbayat2024design}, it provides a crucial benchmark for smaller scenarios, and even under limited runtimes can return high-quality incumbent solutions that serve as meaningful reference points. These methods offer different trade-offs between computational efficiency and solution quality. This variety allows for a broader assessment of the design space.

To provide a clear framework, we first establish the essential sets, parameters, variables, and constraints. This notation serves as the foundation for the exact mathematical formulation, which we construct using ILP.

\subsection{Problem Formulation}
We consider the PON-based fronthaul network design (PON-FD) problem. The architecture connects a set of RUs to DU pools via intermediate optical splitter sites. The optimization determines the active DU pools, the active splitter locations, the number/type of splitters, and the routing topology to minimize TCO under latency and capacity constraints.

\subsubsection*{Sets and Indices:}
\begin{itemize}
    \item $\mathcal{D} = \{1,\dots,N_D\}$: set of candidate DU-pool sites, index $d$.
    \item $\mathcal{S} = \{1,\dots,N_S\}$: set of candidate splitter sites, index $s$.
    \item $\mathcal{R} = \{1,\dots,N_R\}$: set of RUs (ONUs), index $r$.
    \item $\mathcal{T}$: set of splitter types, index $t$, where each level supports $2^t$ RUs.\footnote{In the remainder of this paper, split ratio refers exclusively to the PON splitter fan-out (e.g., 1:8, 1:16), which determines how many RUs share a feeder fiber. This is distinct from the RAN functional split options, which defines how baseband processing is partitioned between the DU and RU (e.g., Option 7.2, Option 6).}
    \item $\mathcal{K}$: Set of DU capacity levels indexed by $k$, where each level supports $2^k$ RUs.

\end{itemize}

\subsubsection*{Parameters:}
\begin{itemize}    
    \item $\delta_{ds}, \delta_{sr}$: Distances (km) connecting candidate DU-pool site $d$ to candidate splitter site $s$, and candidate splitter site $s$ to RU $r$, respectively.\footnote{These distances are pre-calculated for the given set. In this work, we assume Euclidean distances but the formulation is independent of the distance metric choice.}
    \item $\mu_\mathrm{r}$: Service rate of an ONU $r$ in Gb/s 
    \item $C^{\mathrm{df}}, C^{\mathrm{ff}}$: Cost per km of distribution and feeder fiber.
    \item $C^{\mathrm{tr}}$: Trenching cost per km.
    \item $C^{\mathrm{bp}}$: Installation cost for a DU pool.\footnote{This cost covers all additional DU related infrastructure costs for the given technology, including installation and OLT shelf costs.}
    \item $C^{\mathrm{rent}}$: Annual DU pool site rent.
    \item $C^{\mathrm{onu}}_{i}$: Cost of one ONU supporting service rate $i$.
    \item $C^{\mathrm{sp}}_t$: Equipment cost for splitter type $t$.\footnote{This cost includes all additional splitter related equipment costs for the given technology, such as OLT equipment and cabinet costs.}
    \item $C^m$: The average annual maintenance cost ratio relative to the device purchase cost.
    \item $C^{\mathrm{p}}$: Cost of 1~W power consumption per year of network operation, \$/($\mathrm{W}\cdot\mathrm{year}$).

    \item $N_{\mathrm{RU}}^{\max}$: Maximum number of RUs supported by a DU.
    \item $v^{\mathrm{f}}$: Speed of light in optical fiber.
    \item $L^{\mathrm{split}}_t$: Insertion loss (dB) for splitter type $t$.
    \item $L^{\mathrm{fix}}$: Fixed insertion losses excluding the splitter, accounting for losses from connectors, the OLT, and the ONU.
    \item $L^{\mathrm{fib}}$: Fiber attenuation loss (dB/km).
    \item $\Delta L^{\mathrm{m}}$: Safety power loss margin.

    \item $P^{\mathrm{cool}}, P^{\mathrm{du}}, P^{\mathrm{ru}}, P^{\mathrm{onu}}$: Cooling power consumption, DU power consumption per  RU capacity, RU power consumption, and ONU power consumption, respectively.
    \item $L^{\mathrm{budget}}$: Optical power budget (min of Upstream/Downstream).
    \item $T^{\mathrm{FH}}$: Maximum fronthaul latency budget for RUs.
\item $T^{\mathrm{proc}}$: Processing and queueing latency for RUs.
\item $T^{\mathrm{op}}$: Network operation period (in years).

    \item $M$: Large constant (Big-M).
\end{itemize}
\subsubsection*{Decision Variables:}
\begin{itemize}
    \item $f_{dsrt} \in \{0,1\}$: 1 if RU $r$ is served by DU $d$ via splitter site $s$ using splitter type $t$.
    \item $n_{dst} \in \mathbb{Z}_+$: Number of splitters of type $t$ at candidate site $s$ connected to DU $d$.
    \item $z_{ds} \in \{0,1\}$: 1 if a feeder link exists between DU $d$ and candidate splitter site $s$.
    \item $u_d \in \{0,1\}$: 1 if DU pool $d$ is active.
    \item $y_{dk} \in \{0,1\}$: 1 if DU pool $d$ operates at capacity level $k$.
\end{itemize}

\subsubsection*{Objective Function:}
The objective minimizes the TCO over the network operation horizon $T^{\mathrm{op}}$:
\begin{equation}
\min Z \;=\; Z^{\mathrm{CapEx}} \;+\; T^{\mathrm{op}} \cdot Z^{\mathrm{OpEx}}.
\end{equation}

The CapEx term groups all one-time deployment costs:
\begin{align}
Z^{\mathrm{CapEx}} \;=\;& 
\underbrace{
\sum_{d \in \mathcal{D}} \sum_{s \in \mathcal{S}} \sum_{r \in \mathcal{R}} \sum_{t \in \mathcal{T}} (C^{\mathrm{df}} + C^{\mathrm{tr}})\, \delta_{sr}\, f_{dsrt}
}_{\text{Distribution fiber and trenching}} 
\nonumber \\
+\;& 
\underbrace{
\sum_{d \in \mathcal{D}} \sum_{s \in \mathcal{S}} \delta_{ds} \Big( C^{\mathrm{tr}} z_{ds} + C^{\mathrm{ff}} \sum_{t \in \mathcal{T}} n_{dst} \Big)
}_{\text{Feeder trenching and fiber}}
\nonumber \\
+\;& 
\underbrace{
\sum_{d \in \mathcal{D}} C^{\mathrm{bp}} u_d 
+ \sum_{r \in \mathcal{R}} C^{\mathrm{onu}}_{\mu_r}
+ \sum_{d \in \mathcal{D}} \sum_{s \in \mathcal{S}} \sum_{t \in \mathcal{T}} C^{\mathrm{sp}}_t\, n_{dst}
}_{\text{Equipment cost (DU pool, ONU, splitter)}}.
\end{align}

The OpEx term aggregates all recurring annual costs that are then scaled by $T^{\mathrm{op}}$:
\begin{align}
Z^{\mathrm{OpEx}} \;=\;& 
\underbrace{
C^m \Big( \sum_{d \in \mathcal{D}} C^{\mathrm{bp}} u_d 
+ \sum_{r \in \mathcal{R}} C^{\mathrm{onu}}_{\mu_r}
+ \sum_{d \in \mathcal{D}} \sum_{s \in \mathcal{S}} \sum_{t \in \mathcal{T}} C^{\mathrm{sp}}_t\, n_{dst} \Big)
}_{\text{Annual maintenance}}
\nonumber \\
+\;& 
\underbrace{
C^{\mathrm{rent}}\cdot  \sum_{d \in \mathcal{D}} u_d 
}_{\text{Annual site rental}}
\nonumber \\
+\;& 
\underbrace{
C^{\mathrm{p}} \Big( \sum_{d \in \mathcal{D}} \big(P^{\mathrm{cool}} u_d + \sum_{k \in \mathcal{K}} 2^k P^{\mathrm{du}} y_{dk}\big) + N_R (P^{\mathrm{onu}}+P^{\mathrm{ru}}) \Big)
}_{\text{Annual energy}}.
\end{align}
Here, $C^{\mathrm{onu}}_{\mu_r}$ selects the ONU unit cost associated with the service rate $\mu_r$ of RU $r$ from the tabulated cost catalog.

\subsubsection*{Constraints:}

Every RU must be connected via exactly one path:
\begin{equation}
\sum_{d \in \mathcal{D}} \sum_{s \in \mathcal{S}} \sum_{t \in \mathcal{T}} f_{dsrt} = 1, \quad \forall r \in \mathcal{R}.
\end{equation}

The number of RUs assigned to a splitter cannot exceed the number of available splitter ports. Specifically, $2^{t}$ denotes the fan-out capacity (i.e., number of output ports) of a splitter of type $t$:

\begin{equation}
\sum_{r \in \mathcal{R}} f_{dsrt} \;\le\; 2^t \cdot n_{dst}, \quad \forall d \in \mathcal{D}, s \in \mathcal{S}, t \in \mathcal{T}.
\end{equation}

A feeder trench ($z_{ds}$) must exist if any splitters are used for DU $d$ at site $s$:
\begin{equation}
\sum_{t \in \mathcal{T}} n_{dst} \;\le\; M z_{ds}, \quad \forall d \in \mathcal{D}, s \in \mathcal{S}.
\end{equation}

A feeder link requires the DU pool to be active:
\begin{equation}
z_{ds} \;\le\; u_d, \quad \forall d \in \mathcal{D}, s \in \mathcal{S}.
\end{equation}

Propagation delay plus processing delay must be within the fronthaul latency budget:
\begin{equation}
f_{dsrt}\Big(\frac{\delta_{ds} + \delta_{sr}}{v^{\mathrm{f}}} + T^{\mathrm{proc}}\Big) \;\le\; T^{\mathrm{FH}}, \quad \forall d,s,r,t.
\end{equation}

Fiber attenuation plus splitter loss and fixed losses must not exceed the power loss budget:
\begin{align}
f_{dsrt}\Big(L^{\mathrm{fib}}(\delta_{ds} + \delta_{sr}) + L^{\mathrm{split}}_t
+ L^{\mathrm{fix}} + \Delta L^{\mathrm{m}}\Big)\le L^{\mathrm{budget}} 
\quad \forall d,s,r,t.
\end{align}

If a DU is active ($u_d=1$), it must select exactly one capacity level $k$:
\begin{equation}
\sum_{k \in \mathcal{K}} y_{dk} = u_d, \quad \forall d \in \mathcal{D}.
\end{equation}

The total number of RUs 
assigned to DU $d$ must not exceed the chosen capacity:
\begin{equation}
\sum_{s \in \mathcal{S}} \sum_{t \in \mathcal{T}} \sum_{r \in \mathcal{R}} f_{dsrt} \;\le\; \sum_{k \in \mathcal{K}}2^k y_{dk}, \quad \forall d \in \mathcal{D}.
\end{equation}
The total count of RUs served by a DU must not exceed the upper limit:
\begin{equation}
\sum_{k \in \mathcal{K}}2^k y_{dk} \;\le\; N_{\mathrm{RU}}^{\max}, \quad \forall d \in \mathcal{D}.
\end{equation}

\subsection{Alternative Solution Methods}

The ILP formulation described above provides an exact mathematical representation of the problem and guarantees global optimality. NP-hardness follows from the fact that the joint DU placement and splitter assignment sub-problem already contains the uncapacitated facility location problem as a special case, obtained by relaxing the splitter, latency, reach, and power constraints \cite{erbayat2024design}. As the number of RUs or candidate DU locations grows, the solution space expands combinatorially and the branch-and-bound search tree required for a provably optimal solution grows exponentially. This motivates the use of a time-limited ILP that returns a high-quality incumbent solution rather than a proof of optimality for the larger instances considered in this study.

To address this scalability limitation, we investigate three alternative approaches: a meta-heuristic evolutionary strategy, a geometric clustering method, and a graph-based heuristic. These methods trade the guarantee of absolute optimality for the ability to generate near-optimal solutions within reasonable computational timeframes. All three approaches are iterative and are executed until the solution cost converges or a predefined execution time threshold is exceeded. Below, we detail the implementation and modeling principles of each approach.

\subsubsection*{Genetic Algorithm (GA)}
The GA is included as a general-purpose meta-heuristic baseline commonly used in the literature. It explores the fronthaul design space via stochastic population-based search, where candidate solutions encode RU--to--splitter and splitter--to--DU assignments. Physical feasibility and splitter dimensioning are deterministically decoded during fitness evaluation, with constraint violations handled through penalty terms and optional repair steps. Given the large, discrete assignment space and strict physical constraints, part of the search effort is spent evaluating infeasible solutions, particularly in early generations. We adopt a simple GA variant with a fixed parameter configuration across all experiments to ensure fairness and reproducibility, acknowledging that alternative designs and parameter choices may yield different performance trade-offs. Full implementation details are provided in \ifthenelse{\boolean{withappendix}}
{Appendix~\ref{sec:GA}}
{\change{Appendix A of the extended version of this manuscript (arXiv preprint~\cite{arxiv})}}.

\subsubsection*{Improved K-Means Clustering Approach (KMC+)}
The KMC+ method simplifies the topological design problem by treating it as a hierarchical geometric clustering task based on k-means. A known challenge of k-means lies in the need to predefine the number of clusters. KMC+ addresses this issue through an incremental clustering strategy that adaptively increases the number of clusters until all physical constraints are satisfied. The method operates in three stages. First, the algorithm clusters RUs based on distances and maps the resulting centroids to candidate splitter sites. Second, it treats the active splitter locations as weighted demand points and clusters them to determine DU pool locations. Third, it verifies physical layer constraints such as reach, latency, and power budget and performs splitter dimensioning. If any constraint is violated, the algorithm first repeats the search for $I_{\text{inner}}$ iterations using the same cluster configuration. If infeasibility persists, the cluster count corresponding to the failure source (splitter or DU) is increased by a fixed increment, and the procedure restarts.
 This algorithm serves as a geometric baseline because it isolates the impact of distance minimization on total cost and prioritizes geographical proximity over detailed hardware step costs. For a detailed formulation of the KMC+ algorithm, see \ifthenelse{\boolean{withappendix}}{Appendix~\ref{sec:kmeans}}\change{{Appendix B of the extended version of this manuscript (arXiv preprint~\cite{arxiv})}.}

\subsubsection*{Improved Randomized Successive Splitter Assignment (RSSA+)}
The RSSA+ method formulates the fronthaul design problem as a constructive search over a multi-stage directed graph connecting a root node to DUs, splitters, and RUs. The total cost is decomposed into three segments: DU deployment-related costs, splitter deployment-related costs, and RU connection-related costs. The network is constructed iteratively by assigning each RU, in random order, to a feasible DU and splitter that minimizes incremental cost.

Due to the greedy nature of the algorithm, this incremental selection may implicitly prioritize one cost segment over others. To mitigate this bias, an improved variant introduces random weights over the three cost segments. At each iteration, a new set of weights is sampled, the weighted cost is evaluated, and the minimum-cost solution across iterations is retained. This weighting mechanism increases RSSA’s adaptability to a wide range of cost parameter settings by dynamically rebalancing the relative importance of DU, splitter, and RU connection costs across iterations. In addition, feasibility is enforced at every step through explicit checks on reach, latency, optical power budget, and DU capacity, ensuring that all intermediate and final solutions satisfy physical constraints. To that end, we include RSSA+ as a constructive baseline because it reflects practical deployment processes where network elements are added incrementally and cost impacts depend on the existing network state. The complete graph construction and RSSA+ procedure are detailed in \ifthenelse{\boolean{withappendix}}{Appendix~\ref{sec:graph_approach}}\change{{Appendix C of the extended version of this manuscript (arXiv preprint~\cite{arxiv})}.}

%% file: sections/evaluation
\section{Evaluation Methodology}
\label{sec:evaluation}

\begin{table*}[t]
\caption{Benchmarking scenarios and traffic profiles \cite{sup6620205g,ITUTG989,ITUTG9804,ITU_T_G984_2_2019,ITU_T_G987_2_2023_amd1,ITU_T_G9807,ITU_T_G_671}.}
\label{tab:scenarios}
\centering
\begin{tabular}{@{}p{2.2cm}p{3.5cm}p{3.5cm}p{3.5cm}p{3.5cm}@{}}
\toprule
\textbf{Parameter} & \textbf{Scenario 1: \newline Ubiquitous Connectivity} & \textbf{Scenario 2: \newline Hyper-Reliable and Low-Latency Communication} & \textbf{Scenario 3: \newline Immersive Communication} & \textbf{Scenario 4: \newline Massive Communication} \\ \midrule
\textbf{Primary Driver} & Coverage \& Cost Efficiency & Strict Latency \& Reliability & Throughput \& Capacity & Cost/Performance Balance \\ \addlinespace
\textbf{BW (per RU)} & $1$ Gbps & $2$ Gbps & $10$ Gbps & $5$ Gbps \\ \addlinespace
\textbf{Processing Latency ($T^\text{proc}$)} & $300\, \mu s$ & $25\,\mu s$ & $100\,\mu s$ & $100\,\mu s$ \\ \addlinespace
\textbf{Latency Budget ($T_{FH}$)} & $5$ ms & $100\,\mu s$ & $250\,\mu s$ & $150\,\mu s$ \\ \addlinespace
\textbf{Max Splitting Ratio} & High ($1:64$) & Low ($1:8$) & Medium ($1:16$) & High ($1:32$) \\ \addlinespace
\textbf{Map Size} & $20 \times 20$ km$^2$ & $5 \times 5$ km$^2$ & $5 \times 5$ km$^2$ & $10 \times 10$ km$^2$ \\ \addlinespace
\textbf{No. of DUs ($N_D$)} & $\{1, 2, 3, 5, 10\}$ & $\{2, 4, 6, 10, 15\}$ & $\{2, 4, 6, 10, 15\}$ & $\{2, 3, 5, 10\}$ \\ \addlinespace
\textbf{No. of RUs ($N_R$)} & $\{20, 50, 100, 200\}$ & $\{20,50,100,150,300,500\}$ & $\{20,50,100,150,300,500\}$ & $\{50, 100, 200, 300\}$ \\ \addlinespace
\textbf{Splitter Spacing} & $ 2000$ m & $ 150$ m & $ 150$ m & $500$ m \\ \bottomrule
\end{tabular}
\end{table*}

\begin{table}[!ht]
\centering
\caption{Parameters and values based on \cite{ORAN-WG4, 3GPP-38801,sup6620205g,Chen2016JOCN, ranaweera20175g, ranaweera2019optical, musumeci2016optimal, dias2022evolutionary,akhtar2023fronthaul, lisi2017cost, masoudi2020cost, carapellese2015bbu, kokangul2011optimization, fayad2022design, fayad20235g, erbayat2024fronthaul, erbayat2024design, erbayat2025multi, erbayat2025toward,erbayat2025latency, erbayat2025design,arevalo2017optimization,FBA_Cartesian_2025,iea_energy_prices, ITU_T_G_671,ITU_T_G9807,ITU_T_G984_2_2019,ITU_T_G987_2_2023_amd1}.}
\label{tab:parameters}
\setlength{\tabcolsep}{3pt}
\renewcommand{\arraystretch}{1.2} 
\begin{tabular}{@{}ll | ll@{}}
\hline
\textbf{Parameter} & \textbf{Value} & \textbf{Parameter} & \textbf{Value} \\
\hline
$C^{\mathrm{df}}$ & 2000\$/km & $N_{\mathrm{RU}}^{\max}$ & 64 \\
$C^{\mathrm{ff}}$ & 3000\$/km & $v^{\mathrm{f}}$ & $2\cdot 10^8$m/s \\
$C^{\mathrm{tr}}$ & 16000\$/km& $L^{\mathrm{fib}}$ & 0.25dB/km \\
$C^{\mathrm{bp}}$ & 135000\$ & $L^{\mathrm{fix}}$ & 3dB \\
$C^{\mathrm{rent}}$ & 10000\$/yr & $\Delta L^{\mathrm{m}}$ & 2dB \\
$C^m$ & 10\% & $L^{\mathrm{budget}}$ & 32dB \\
$C^{\mathrm{p}}$ & 1.5\$/W$\cdot$yr & $P^{\mathrm{cool}}$ &  500W \\
$T^{\mathrm{op}}$ & 20yrs & $P^{\mathrm{du}}$ & 100W \\
$P^{\mathrm{ru}}$ & 60W & $P^{\mathrm{onu}}$ & 4W \\
\hline

\multicolumn{4}{l}{$C^{\mathrm{onu}}_{i}$: $\{100, 200, 400\}$ for rates of $\{1, 2.5, 10\}$~Gbps} \\
\multicolumn{4}{l}{ $C^{\mathrm{sp}}_t$: (5000 + $t\cdot70)\$$ for a $2^{t}$ splitter}  \\
\multicolumn{4}{l}{$L^{\mathrm{split}}_{t}$:  $t\cdot3.5\mathrm{dB}$ for a $2^{t}$ splitter} \\
\hline
\end{tabular}
\end{table}

To evaluate the performance of the proposed benchmarking framework and optimization algorithms, we utilize a Python-based simulation environment. This environment models the physical and logical constraints of PON-based fronthaul networks.

We define the simulation area based on the deployment scenario and randomly place candidate DU pools and RUs using a uniform spatial distribution. For benchmarking purposes, DUs and RUs are assumed to be homogeneous, with identical DU capacities and costs, as well as uniform RU processing latency and traffic demand. This simplification enables fair comparison across algorithms while preserving a broadly applicable benchmarking environment, since the proposed methods can readily accommodate heterogeneous parameters. Candidate splitter locations are sampled periodically and serve as potential aggregation points that the optimization algorithms may select and activate depending on the scenario configuration.

Then, we define four distinct evaluation scenarios to ensure the proposed framework is robust across a wide variety of deployment conditions. These scenarios vary by user density, traffic composition, and physical constraints, inspired by the IMT-2030 usage framework, which captures the evolution of mobile services toward immersive applications, mission-critical communication, large-scale machine connectivity, and ubiquitous coverage \cite{ITURM2516}. Rather than modeling individual applications, the selected scenarios abstract the dominant operating regimes expected in next-generation networks and translate them into representative fronthaul design constraints \cite{ITURM2160}. The four scenarios are described below.

\begin{itemize}

\item \textbf{Scenario 1: Ubiquitous Connectivity:}  
This scenario reflects IMT-2030's focus on extending connectivity to rural and sparsely populated areas. Traffic is dominated by low-rate and delay-tolerant services, while long fiber distances make optical power budget and reach the primary constraints. Higher-layer functional splits and large PON split ratios enable cost-efficient coverage over wide geographic areas.

\item \textbf{Scenario 2: Hyper-Reliable and Low-Latency Communication:}  
This scenario represents mission-critical urban deployments such as industrial automation and autonomous transport. Stringent reliability and sub-millisecond latency requirements impose hard fronthaul delay constraints, favoring low-layer functional splits and short DU--RU separation. Network design is therefore driven by latency feasibility rather than cost minimization.

\item \textbf{Scenario 3: Immersive Communication:}  
This scenario captures dense urban environments supporting immersive XR and high-capacity multimedia services. Aggregate throughput demand dominates fronthaul design, while latency constraints are moderate. High-bandwidth PON technologies and carefully chosen split ratios are required to avoid congestion under peak load.

\item \textbf{Scenario 4: Massive Communication:}  
This scenario reflects mixed suburban environments with large numbers of connected devices and moderate broadband demand. No single constraint dominates; instead, fronthaul design balances cost, capacity, and latency using standard functional splits and intermediate PON configurations.

\end{itemize}

Table~\ref{tab:scenarios} summarizes the key requirements and characteristics associated with each evaluation scenario. Each scenario is defined by target latency and bandwidth requirements, which capture the dominant service constraints expected in that operating regime. For benchmarking purposes in fronthaul network design, we do not explicitly model fine-grained bandwidth allocation or traffic scheduling. Instead, bandwidth demand, traffic characteristics, and latency requirements are reflected implicitly through their impact on processing latency and allowable PON split ratios. In particular, higher split ratios increase both queuing delay and aggregate bandwidth demand on shared feeder links. Consequently, for each scenario, we impose a maximum admissible split ratio that captures the trade-off between infrastructure sharing, latency feasibility, and capacity constraints.

To rigorously evaluate the proposed algorithms under these contrasting conditions, we establish a unified simulation environment. Although cost and technical parameters naturally vary across deployments, meaningful benchmarking requires a standardized and controlled framework. We therefore adopt broad industry-average values, rather than isolated case studies, to define a representative baseline applicable to typical deployments. Table~\ref{tab:parameters} summarizes the resulting framework, including the power consumption models, cost catalog, and optical link budget parameters used consistently across all scenarios.

%% file: sections/results
\begin{figure*}[t]
    \centering
    \caption{Fronthaul performance across deployment scenarios. In every subplot, the vertical axis reports total deployment cost in units of $10^6$ USD. The horizontal axis shows either the number of RUs or the number of candidate DU locations as indicated in each sub-caption. Bars are stacked by cost component (trenching, fiber, equipment, maintenance, site rent, and energy) and grouped by algorithm (ILP, GA, KMC+, RSSA+), with error bars denoting $95\%$ confidence intervals over the 25 independently generated topologies. The legend identifying the cost components and the algorithms is shown in the bottom-right panel.}
    \label{fig:all_scenarios}

    \begin{subfigure}{0.32\textwidth}
        \includegraphics[width=\linewidth]{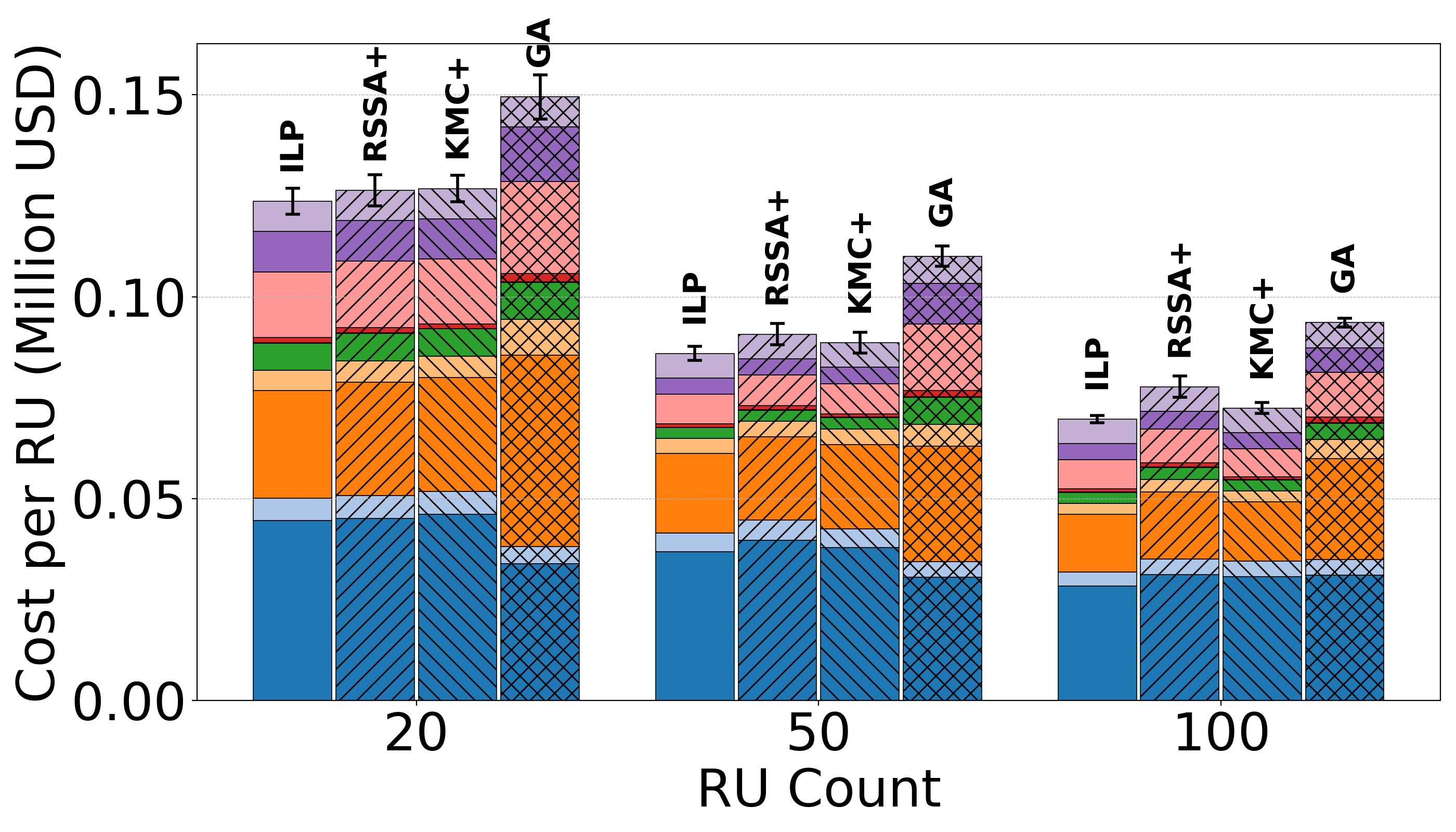}
        \caption{Scenario 1: RU count versus total cost for 3 candidate DU positions. }
    \end{subfigure}
    \begin{subfigure}{0.32\textwidth}
        \includegraphics[width=\linewidth]{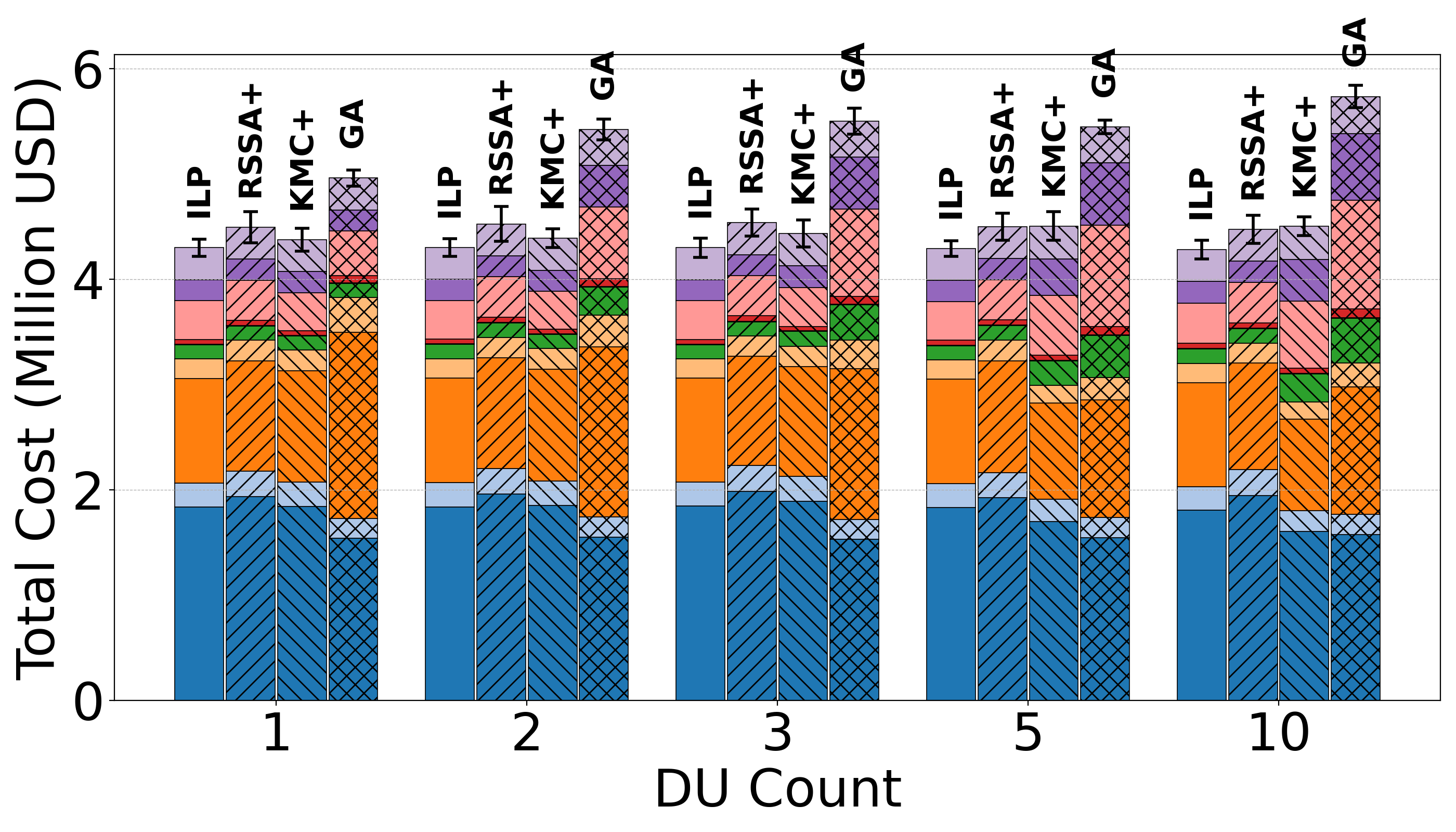}
        \caption{Scenario 1: DU count versus total cost for 50 RUs.}
    \end{subfigure}
    \begin{subfigure}{0.32\textwidth}
        \includegraphics[width=\linewidth]{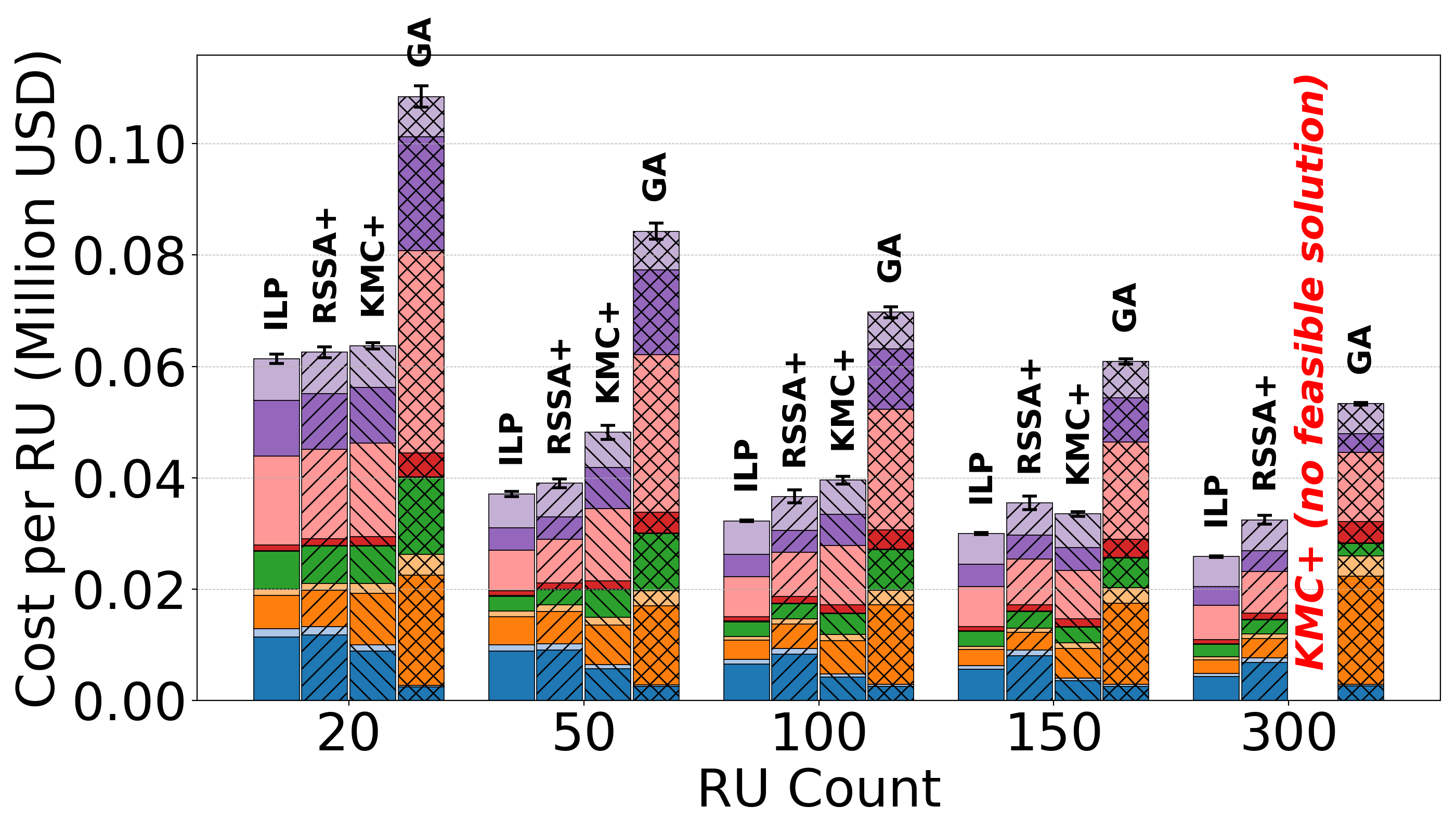}
        \caption{Scenario 2: RU count versus total cost for 6 candidate DU positions.}
    \end{subfigure}

    \medskip

    \begin{subfigure}{0.32\textwidth}
        \includegraphics[width=\linewidth]{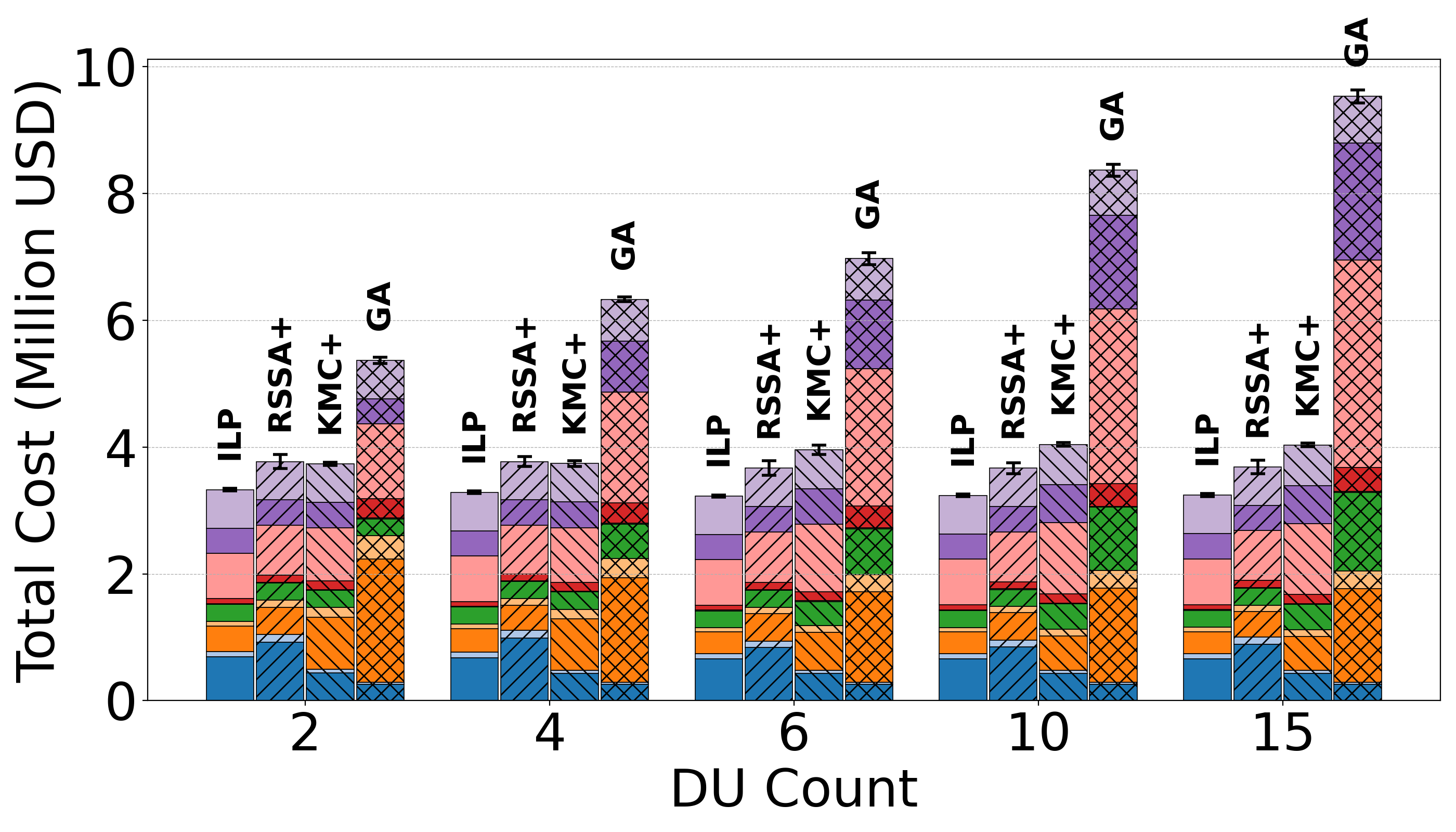}
        \caption{Scenario 2: DU count versus total cost for 100 RUs.}
    \end{subfigure}
    \begin{subfigure}{0.32\textwidth}
        \includegraphics[width=\linewidth]{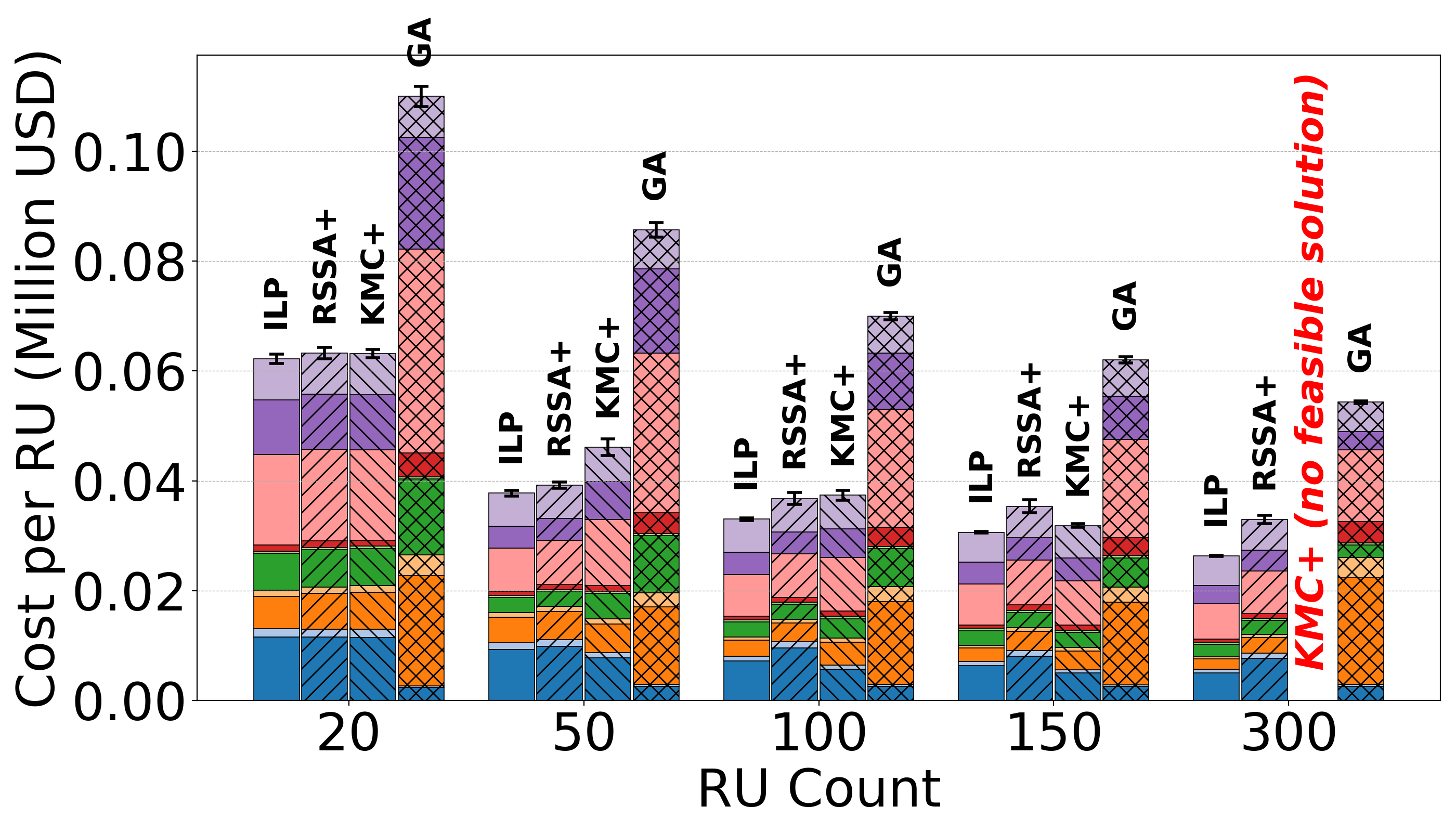}
        \caption{Scenario 3: RU count versus total cost for 6 candidate DU positions.}
    \end{subfigure}
    \begin{subfigure}{0.32\textwidth}
        \includegraphics[width=\linewidth]{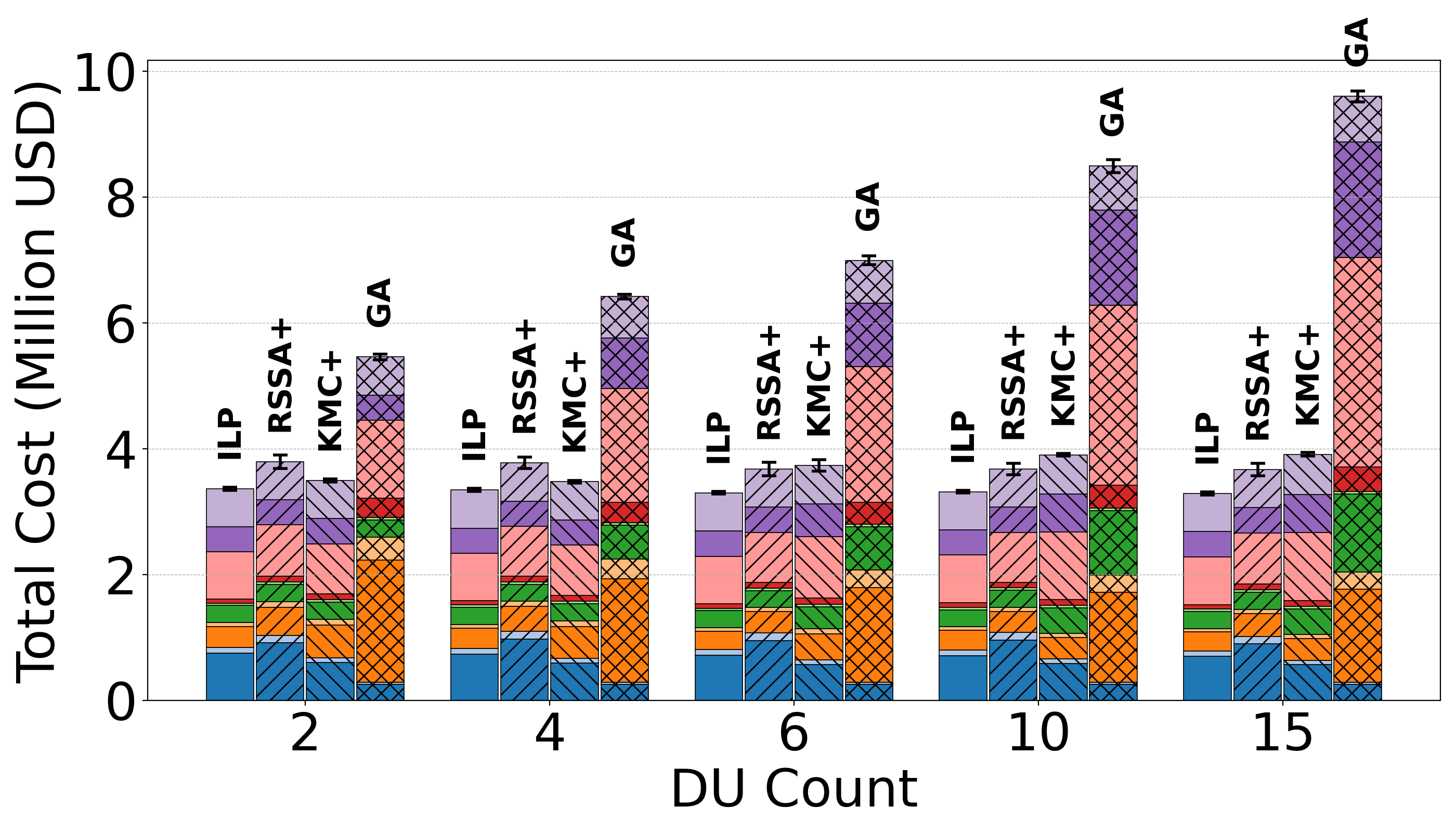}
        \caption{Scenario 3: DU count versus total cost for 100 RUs.}
    \end{subfigure}

    \medskip

    \begin{subfigure}{0.32\textwidth}
        \includegraphics[width=\linewidth]{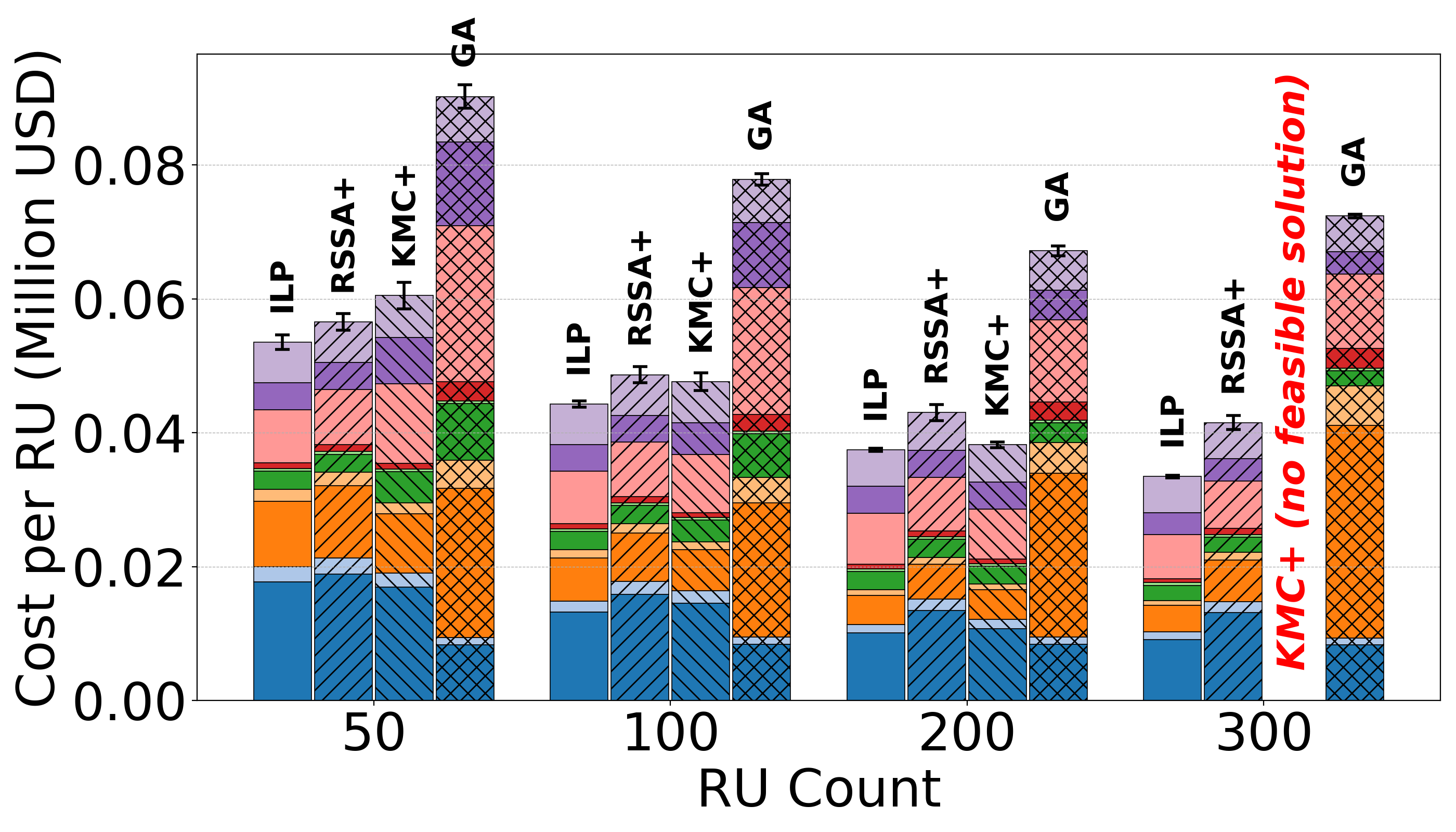}
        \caption{Scenario 4: RU count versus total cost for 5 candidate DU positions.}
    \end{subfigure}
    \begin{subfigure}{0.32\textwidth}
        \includegraphics[width=\linewidth]{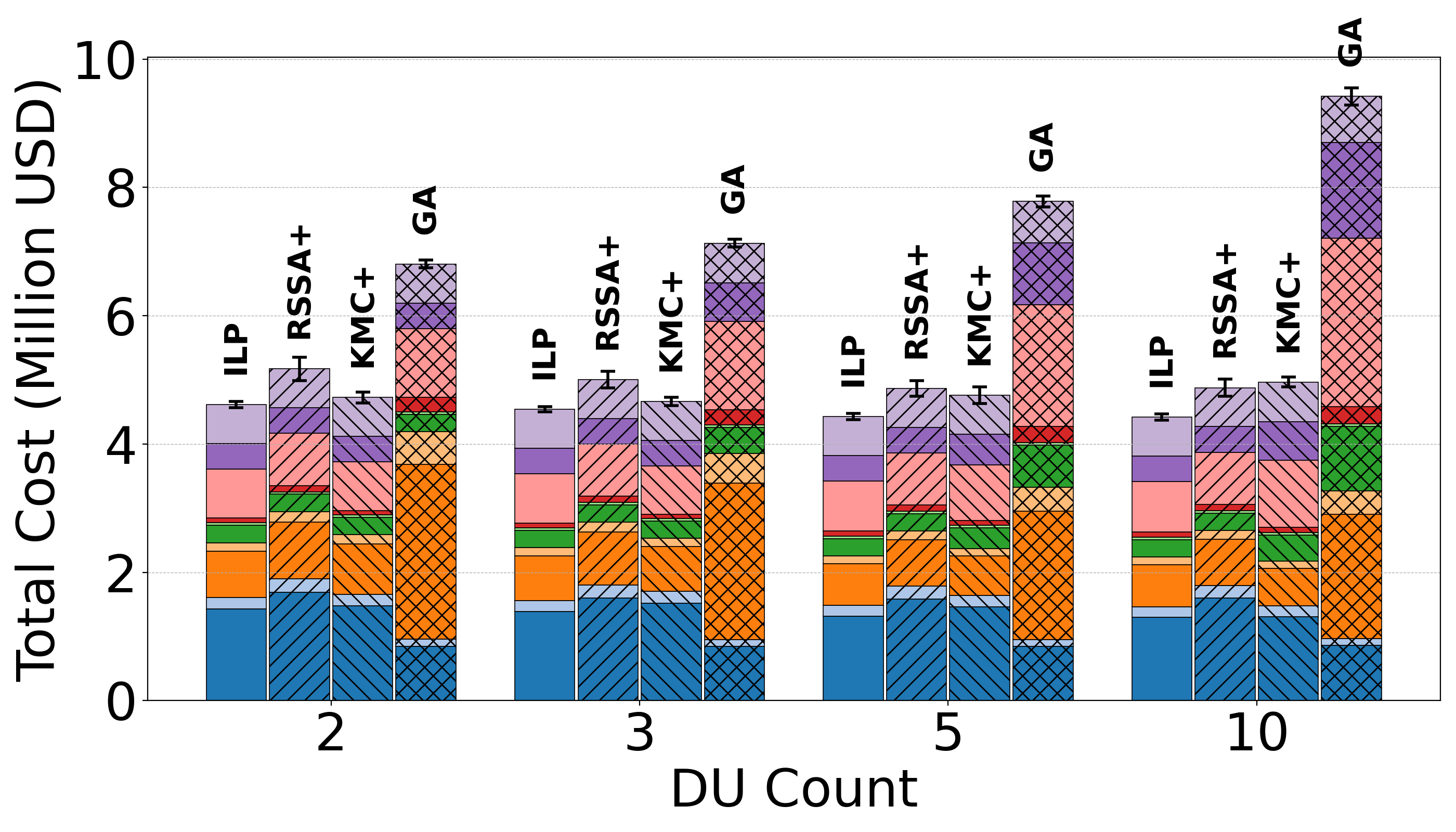}
        \caption{Scenario 4: DU count versus total cost for 100 RUs.}
    \end{subfigure}
    \begin{subfigure}{0.32\textwidth}
        \includegraphics[width=\linewidth]{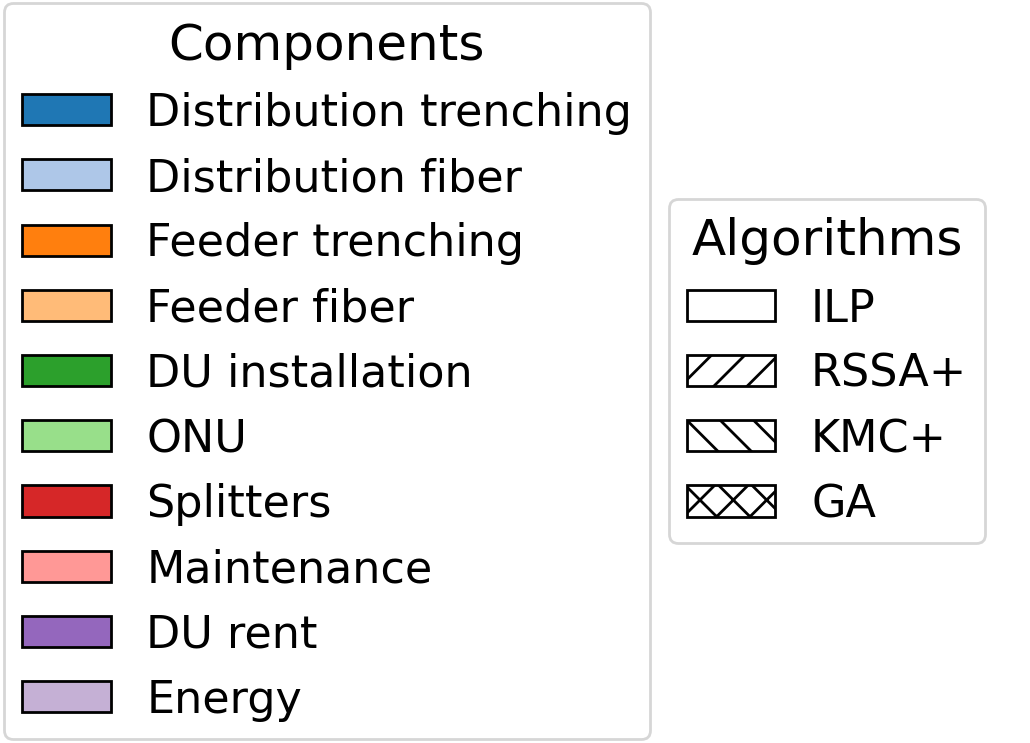}
        \caption{Legend}
    \end{subfigure}
\end{figure*}
\section{Numerical Results}
\label{sec:numerical_results}

We now evaluate the performance of the proposed optimization approaches under the standardized benchmarking framework. Figure~\ref{fig:all_scenarios} summarizes the performance of the four optimization approaches—ILP, GA, KMC+, RSSA+. For each scenario, we evaluate the total deployment cost as a function of both the number of RUs and the number of candidate DU locations. We use the unified cost and constraint model described in Tables~\ref{tab:scenarios} and~\ref{tab:parameters}. Cost breakdowns include trenching, fiber, equipment, maintenance, site rent, and energy components. Results are averaged over 25 independently generated network topologies, and confidence intervals are shown above the bars. RU positions are sampled uniformly at random over the map area, and DU positions are also sampled uniformly; however, for each $(N_{\mathrm{D}}, N_{\mathrm{R}})$ pair, DU locations are held fixed across all topologies to ensure consistent comparison. In addition, each heuristic is executed with a maximum runtime of $1000$ seconds or until no improvement in total cost is observed over the last $1200$ iterations, whichever occurs first, while the ILP is solved with a time limit of $3600$ seconds and its best incumbent solution is recorded if optimality is not reached.\footnote{Across all scenarios in Figure \ref{fig:all_scenarios}, the ILP completes within the time limit in 23 instances for $(N_D, N_R) = (3, 50)$ and in 4 instances for $(3, 100)$ under Scenario 1.} The complete benchmarking framework and datasets are publicly shared in \cite{technicalNote}.

\begin{figure}[t]
    \centering
    \begin{subfigure}{0.48\linewidth}
        \includegraphics[width=\linewidth]{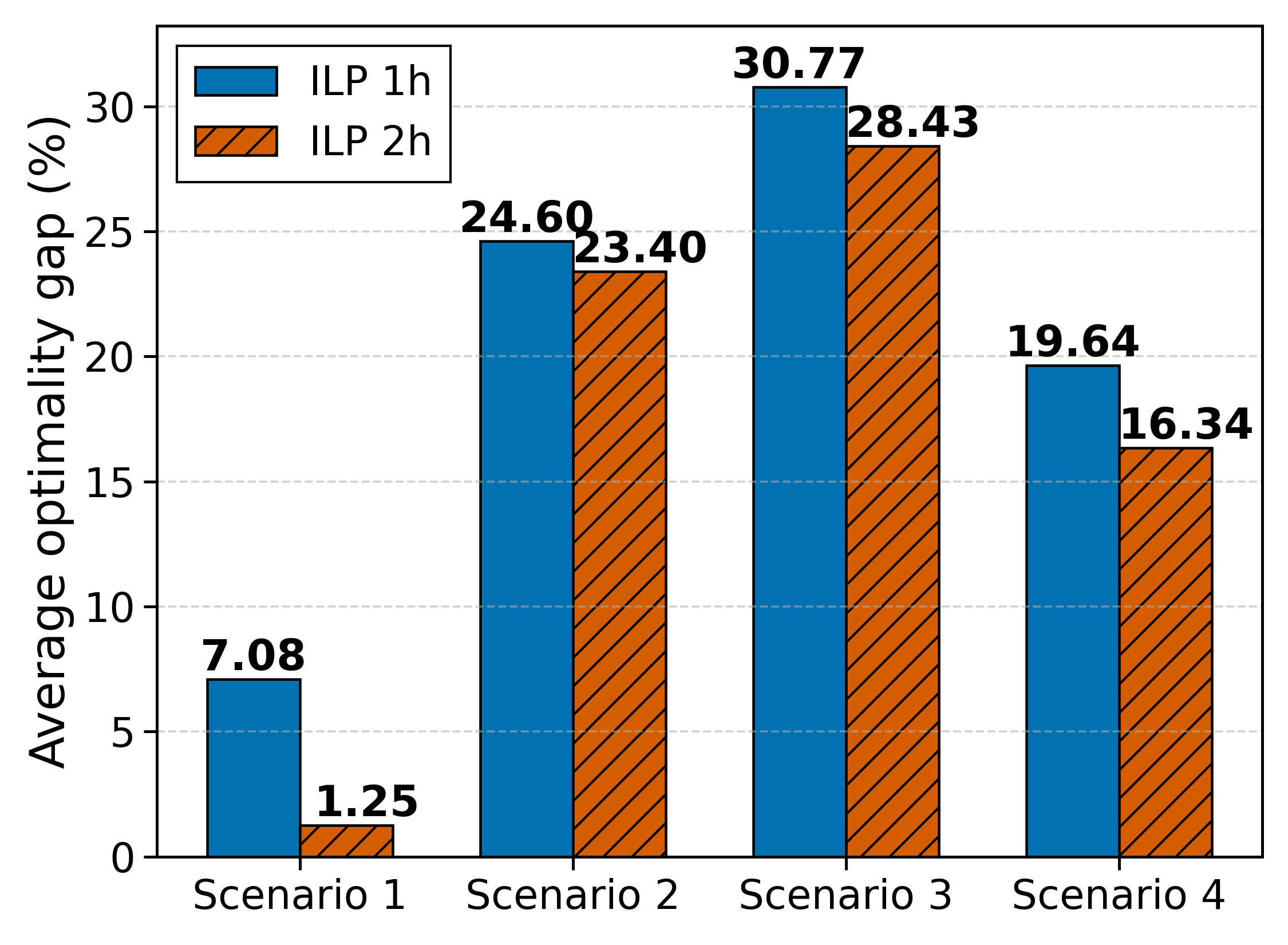}
        \caption{Average MIP gap per scenario at the $1$\,h and $2$\,h budgets.}
    \end{subfigure}
    \hfill
    \begin{subfigure}{0.48\linewidth}
        \includegraphics[width=\linewidth]{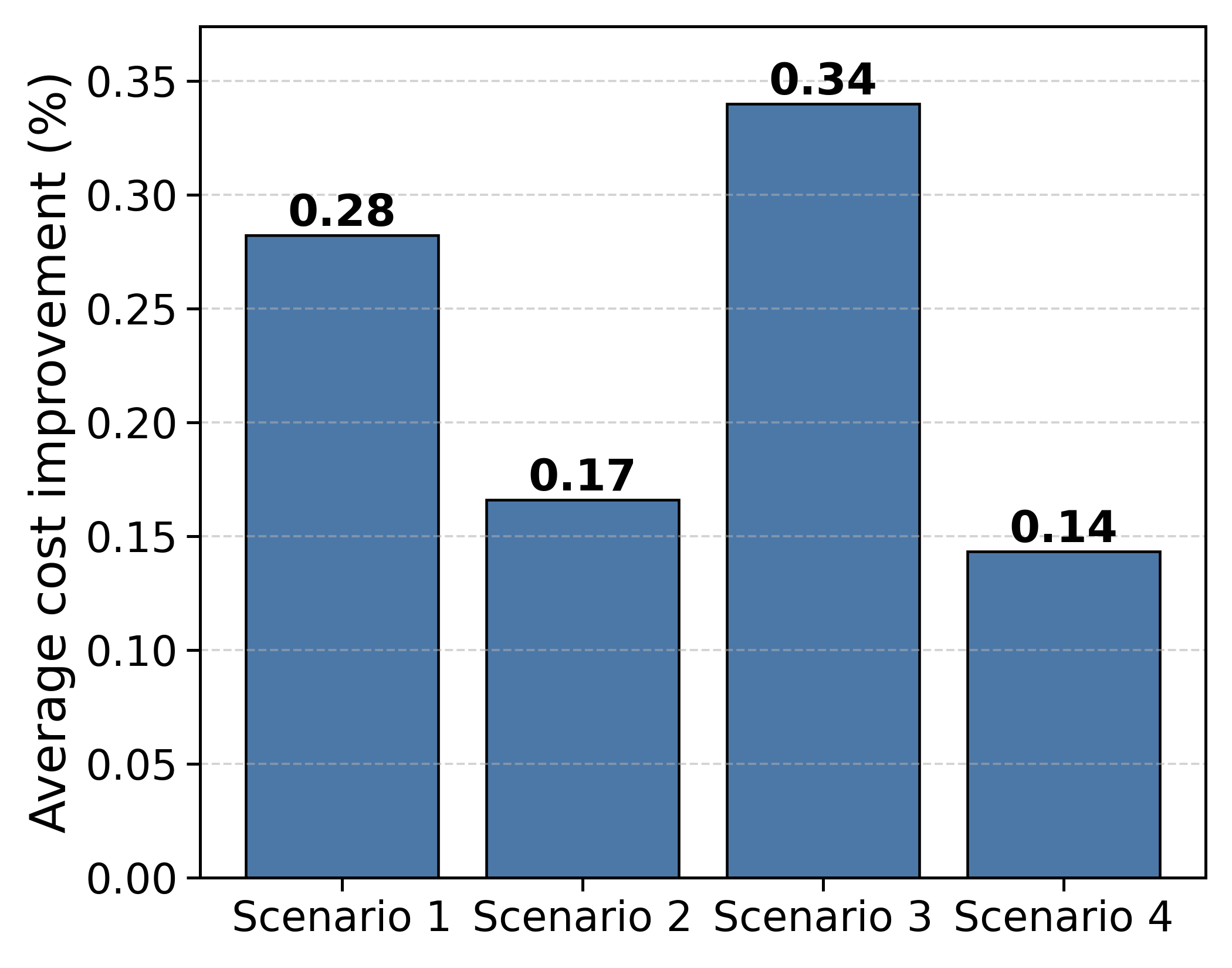}
        \caption{Average total cost improvement from $1$\,h to $2$\,h.}
    \end{subfigure}
    \caption{Runtime versus optimality trade-off of the time-limited ILP.}
    \label{fig:ilp_runtime_quality}
\end{figure}

Figures~\ref{fig:all_scenarios}(a) and~\ref{fig:all_scenarios}(b) present results for Scenario~1, which represents wide-area deployments with relaxed latency constraints and high splitting ratios. In this regime, trenching and fiber costs dominate total deployment cost due to long DU--RU distances. As a result, Scenario~1 exhibits the highest cost per RU across all scenarios. Increasing the number of candidate DU locations reduces feeder distances and lowers total cost up to a point, after which additional DU site costs introduce a clear trade-off.

Figures~\ref{fig:all_scenarios}(c) and~\ref{fig:all_scenarios}(d) correspond to Scenario~2, where strict latency constraints and low split ratios lead to dense deployments. In this case, device-related costs, maintenance, site rent, and energy expenditures increase significantly due to higher DU and splitter activation. Feasible solutions do not exist for all RU--DU pairs, and KMC+ fails to produce feasible solutions at high RU counts, particularly for $N_R = 300$. This behavior arises because KMC+ prioritizes distance minimization without explicit constraint handling, whereas RSSA+ enforces feasibility throughout the construction process.

Results for Scenario~3 are shown in Figs.~\ref{fig:all_scenarios}(e) and~\ref{fig:all_scenarios}(f). Similar to Scenario~2, high RU density and elevated bandwidth demand increase device-related and operational costs. KMC+ again becomes infeasible at large RU counts, while RSSA+ remains close to the ILP baseline across all feasible configurations. GA incurs higher costs overall.

Figures~\ref{fig:all_scenarios}(g) and~\ref{fig:all_scenarios}(h) present Scenario~4, which exhibits behavior between Scenarios~1 and~2-3. Both civil works and device-related costs contribute materially to total cost. Feasible solutions exist across most configurations, and RSSA+ outperforms KMC+ and GA while remaining close to the ILP solution.

Across all scenarios, increasing the number of DUs reduces deployment cost by shortening feeder distances, while excessive DU deployment increases site-related costs. ILP and RSSA+ capture this trade-off through selective DU activation, whereas KMC+ relies on fixed cluster counts and tends to activate additional DU locations as DU availability increases, which leads to higher site-related costs. Consequently, KMC+ performs well in dense RU deployments, where spatial clustering supports efficient infrastructure sharing, but it exhibits a higher risk of infeasible solutions as RU density approaches splitter or DU capacity limits due to limited adaptability, as shown in Scenario 2-3. RSSA+, by contrast, constructs solutions incrementally at the RU level and enforces feasibility at each step, which makes it robust to capacity saturation and effective when multiple DU options are available. This RU-by-RU construction also makes RSSA+ naturally extensible to heterogeneous latency and processing requirements, whereas KMC+ assumes homogeneous nodes within clusters, although heterogeneity is not explored in this study to preserve a controlled benchmarking environment. Overall, KMC+ and RSSA+ closely track the ILP solution across most configurations, whereas the simple GA implementation performs poorly because evolutionary search struggles to maintain feasibility under strict latency, capacity, and power constraints.

Finally, although the ILP formulation is computationally intractable for most problem sizes considered, it remains a strong reference point when solved under a time limit. The $3600$-second budget used in Figure~\ref{fig:all_scenarios} is a pragmatic default for the benchmarking scale rather than a solver limit, since each $(N_D, N_R)$ pair is evaluated over $25$ independent topologies across four scenarios for a total of more than $900$ ILP instances. A multi-hour per-instance budget is therefore infeasible at this aggregate scale, while a single planning instance can be pushed arbitrarily further on the released code under any user-defined time limit. Figure~\ref{fig:ilp_runtime_quality} quantifies the resulting runtime versus optimality trade-off. Scenario~1 closes from a $7.08\%$ average MIP gap at $1$\,h to $1.25\%$ at $2$\,h, while the dense and latency-bound Scenarios~2 and~3 retain residual gaps of $23.40\%$ and $28.43\%$ even at $2$\,h. Crucially, the cost benefit of this additional hour is very low. The average total cost improves by less than $0.35\%$ in every scenario when the budget is doubled, and by less than $0.2\%$ in the hardest ones, so the residual gap in the harder scenarios reflects a slack LP relaxation rather than recoverable cost. The benchmarking conclusions in Figure~\ref{fig:all_scenarios} are therefore robust to the per-instance budget. Even under a suboptimal termination, the ILP incumbent consistently outperforms all heuristic methods and provides a meaningful lower bound on achievable cost. While exact ILP-based optimization is often overlooked in large-scale network planning, these results show that time-limited ILP at $3600$\,s remains a valuable and informative baseline.\footnote{We note that the synthetic spatial layouts used here are chosen to isolate algorithmic behavior under controlled scenario conditions. The benchmarking framework, however, accepts arbitrary geographic coordinates for candidate DU sites, splitter sites, and RU locations. Plugging in real urban topologies such as a municipal cell site database is therefore a direct extension of the same framework and does not require any structural change to the formulation or the algorithms. A dedicated evaluation of the proposed methods on real urban datasets and a comparison with the synthetic layouts of this paper is planned as a natural next step.}

%% file: sections/conclusion
\section{Conclusion}\label{s:Conclusion}
In this work, we examined the transition of mobile networks toward 5G and 6G RAN architectures, identifying PONs as a critical solution for cost-effective fronthaul transport. We addressed the lack of standardized evaluation models in current literature, which prevents the objective comparison of diverse optimization strategies, by proposing a unified benchmarking framework that standardizes cost catalogs and deployment scenarios. Furthermore, we formulated the network design problem using ILP to establish optimality bounds and evaluated three scalable heuristic strategies: GA, KMC+, RSSA+. Simulation results demonstrate that time-limited ILP, despite terminating with suboptimal solutions, provides a strong and informative baseline that outperforms standard heuristics and is often overlooked in existing fronthaul planning literature. Among scalable methods, RSSA+ consistently achieves near-ILP performance while guaranteeing feasibility across all scenarios, which highlights the need for structured and constraint-aware algorithms beyond simple heuristic designs.

%% file: sections/appendix
This appendix provides detailed descriptions of the heuristic algorithms used in this study. We first define a set of common additional parameters that are shared across all heuristic methods.

Let $T_{\text{run}}$ denote the run time limit in seconds, and let $I_{\text{pat}}$ denote the early-stopping patience counter, such that an algorithm terminates after $I_{\text{pat}}$ consecutive iterations without sufficient improvement. We denote by $\epsilon$ the improvement threshold, such that only cost reductions larger than $\epsilon$ reset the patience counter. We set $\epsilon = 10^{-3}$, $I_{\text{pat}} = 1200$, and $T_{\text{run}} = 1000$ seconds for all heuristic algorithms to ensure a fair and consistent computational budget across methods.

\section{Genetic Algorithm (GA)}
\label{sec:GA}

We employ a simple yet scalable GA baseline to explore the combinatorial fronthaul design space. The GA evolves a population of candidate topologies, where each individual represents a two-tier assignment (RU$\rightarrow$Splitter and Splitter$\rightarrow$DU). Constraint satisfaction (latency and optical power budget) and splitter configuration decisions are not explicitly encoded in the chromosome; instead, they are deterministically decoded during fitness evaluation. Infeasible solutions are discouraged using a combination of hard penalties (dominating the objective) and soft penalties (proportional to violation magnitudes), and a lightweight repair operator is optionally applied to offspring.

The GA is executed under a hard runtime budget $T_\text{run}$. The implementation checks the elapsed time inside the deepest evaluation and reproduction loops and immediately returns the best solution found so far when the time limit is exceeded. Convergence is controlled using an improvement threshold $\varepsilon$ and a patience parameter $I_\text{pat}$, both supplied externally.

\subsection{Chromosome Encoding and Representation}

We adopt an integer encoding that jointly captures both tiers of the fronthaul topology. A chromosome is
\begin{equation}
\mathcal{X} = (\mathbf{s}, \mathbf{d}),
\end{equation}
where
\begin{itemize}
    \item $\mathbf{s} = [s_1, s_2, \dots, s_{N_R}]$, with $s_r \in \{0,\dots,N_S-1\}$, assigns each RU $r$ to a candidate splitter site;
    \item $\mathbf{d} = [d_1, d_2, \dots, d_{N_S}]$, with $d_s \in \{0,\dots,N_D-1\}$, assigns each splitter site $s$ to a DU.
\end{itemize}

A splitter site is considered {active} if at least one RU is assigned to it; splitter-to-DU decisions for inactive sites do not affect evaluation. This representation reduces dimensionality relative to the ILP decision tensor $f_{dsrt}$, while retaining flexibility to explore diverse hierarchical topologies.

Importantly, splitter types and splitting ratios are not directly encoded. Instead, for each active splitter, the algorithm infers (i) a feasible splitting depth (if any exists under the optical budget) and (ii) the number of physical splitters required to serve its RU load.

\subsection{Fitness Evaluation and Constraint Handling}

The GA minimizes a composite objective
\begin{equation}
J(\mathcal{X}) = \Psi(\mathcal{X}) + \Omega(\mathcal{X}),
\end{equation}
where $\Psi(\mathcal{X})$ is a proxy for the TCO and $\Omega(\mathcal{X})$ penalizes constraint violations.

The cost proxy $\Psi(\mathcal{X})$ aggregates:
(i) distribution fiber and trenching costs proportional to RU--splitter distances,
(ii) feeder trenching and fiber costs proportional to splitter--DU distances and the number of physical splitters,
(iii) splitter, DU, and ONU equipment costs,
(iv) maintenance, site rent, and energy costs over the operational horizon, and
(v) a regularization term proportional to the number of active splitter sites.

Constraint violations are handled through
\begin{equation}
\Omega(\mathcal{X}) \;=\; 10^{9} \cdot V_{\text{hard}} \;+\; 10^{6} \cdot \Big(
\sum_s \delta^{\text{reach}}_s +
\sum_s \delta^{\text{lat}}_s +
\sum_d \delta^{\text{cap}}_d
\Big),
\end{equation}
where $V_{\text{hard}}$ counts hard violations, namely:
(i) splitter reach violations,
(ii) latency violations,
(iii) absence of any feasible splitter type under the optical budget, and
(iv) DU capacity violations.
The soft terms $\delta^{\text{reach}}_s$, $\delta^{\text{lat}}_s$, and $\delta^{\text{cap}}_d$ quantify the magnitude of reach, latency, and DU overload violations, respectively. This formulation strongly discourages infeasible solutions while allowing near-feasible individuals to guide the search. When available, the best solutions identified by the GA are verified using a canonical feasibility checker and rescored using the exact cost function. If a solution is feasible under the canonical model, its canonical cost replaces the proxy fitness for best-solution tracking.

\subsection{Initialization, Operators, Repair, and Early Stopping}

Prior to the GA loop, nearest-neighbor candidate sets are computed using precomputed distance matrices:
each RU selects uniformly among its $k_{nn}=8$ nearest splitters, and each splitter selects uniformly among its $k_{nn}=8$ nearest DUs. This biases the initial population toward geographically plausible solutions while maintaining diversity. On the other hand, parent selection uses $k$-way tournament selection with $k=3$. Uniform crossover is applied independently to RU--splitter and splitter--DU assignments with probability $p_c = 0.85$. Mutation is applied independently as follows:
\begin{itemize}
    \item RU reassignment with probability $p_m^{RU} = 0.03$, selecting a new splitter from the RU's nearest-neighbor set;
    \item Splitter-to-DU reassignment with probability $p_m^{SD} = 0.02$, selecting a new DU from the splitter's nearest-neighbor set.
\end{itemize}
An elitist strategy preserves the top $10\%$ of individuals unchanged in each generation.

Afterwards, with probability $p_{\text{rep}} = 0.40$, offspring are subjected to a lightweight repair procedure. The repair sequentially addresses infeasibility by:
(i) reassigning splitters with no feasible optical type to their nearest DU,
(ii) rerouting a small fraction of RUs away from splitters violating reach or latency constraints, and
(iii) migrating lightly loaded splitters from overloaded DUs to alternative nearby DUs with remaining capacity.
All repair steps are time-aware and terminate immediately if the runtime limit is exceeded.

Finally, a new solution is considered an improvement only if it reduces the best-known objective by at least $\varepsilon$. Generation-level evolution terminates early if no improvement is observed for $I_{\text{pat}}$ consecutive generations. In addition, the GA operates under a strict runtime budget $T_{\text{run}}$. Elapsed time is checked inside the innermost evaluation and reproduction loops. If the time limit is exceeded at any point, the algorithm immediately returns the best solution found so far, regardless of feasibility. The full GA process is summarized in Algorithm \ref{alg:GA_detailed}

\begin{algorithm}[!ht]
\caption{Genetic Algorithm (GA)}
\label{alg:GA_detailed}
\begin{algorithmic}[1]
\Require Sets $\mathcal{R}$, $\mathcal{S}$, $\mathcal{D}$;
runtime budget $T_{\text{run}}$; improvement threshold $\varepsilon$; patience $I_{\text{pat}}$;
GA parameters $N_{pop}$, $N_{gen}$, $e$, $k$, $p_c$, $p_m^{RU}$, $p_m^{SD}$, $p_{\text{rep}}$
\Ensure Best fronthaul topology $\mathcal{X}_{best}$ found within $T_{\text{run}}$

\State Precompute RU--Splitter and Splitter--DU nearest-neighbor candidate sets
\State $t_0 \leftarrow$ current time
\State Initialize $\mathcal{X}_{best}$ with $J(\mathcal{X}_{best}) \leftarrow +\infty$
\State $r_{\text{no-imp}} \leftarrow 0$

\While{True}
    \If{current time $- t_0 > T_{\text{run}}$}
        \State \Return $\mathcal{X}_{best}$
    \EndIf

    \State Initialize population $P$ using distance-aware random assignment
    \State Optionally seed a fraction of $P$ from a feasible heuristic solution
    \State $g_{\text{no-imp}} \leftarrow 0$, $J^{\text{restart}}_{\text{best}} \leftarrow +\infty$

    \For{$g = 1$ to $N_{gen}$}
        \If{current time $- t_0 > T_{\text{run}}$}
            \State \Return $\mathcal{X}_{best}$
        \EndIf

        \State Evaluate $J(\mathcal{X})$ for all $\mathcal{X} \in P$
        \State $\mathcal{X}^\star \leftarrow \arg\min_{\mathcal{X} \in P} J(\mathcal{X})$

        \If{$J(\mathcal{X}^\star) < J^{\text{restart}}_{\text{best}} - \varepsilon$}
            \State $J^{\text{restart}}_{\text{best}} \leftarrow J(\mathcal{X}^\star)$
            \State $g_{\text{no-imp}} \leftarrow 0$
        \Else
            \State $g_{\text{no-imp}} \leftarrow g_{\text{no-imp}} + 1$
        \EndIf

        \State Update global best $\mathcal{X}_{best}$ if improved by at least $\varepsilon$
        \If{$g_{\text{no-imp}} \ge I_{\text{pat}}$}
            \State \textbf{break} \Comment{generation-level early stopping}
        \EndIf

        \State Preserve top $e$ fraction of elite individuals in $P_{\text{next}}$
        \While{$|P_{\text{next}}| < N_{pop}$}
            \If{current time $- t_0 > T_{\text{run}}$}
                \State \Return $\mathcal{X}_{best}$
            \EndIf
            \State Select parents using $k$-way tournament selection
            \State Apply uniform crossover with probability $p_c$
            \State Apply mutation to RU and splitter--DU genes with probabilities $p_m^{RU}$ and $p_m^{SD}$
            \State With probability $p_{\text{rep}}$, apply repair heuristic
            \State Add offspring to $P_{\text{next}}$
        \EndWhile

        \State $P \leftarrow P_{\text{next}}$
    \EndFor

    \If{global best improved by at least $\varepsilon$}
        \State $r_{\text{no-imp}} \leftarrow 0$
    \Else
        \State $r_{\text{no-imp}} \leftarrow r_{\text{no-imp}} + 1$
    \EndIf

    \If{$r_{\text{no-imp}} \ge I_{\text{pat}}$}
        \State \textbf{break} \Comment{restart-level early stopping}
    \EndIf
\EndWhile

\State \Return $\mathcal{X}_{best}$
\end{algorithmic}
\end{algorithm}

\section{Improved K-Means Clustering (KMC+)}\label{sec:kmeans}

To address the dependency between network topology and physical layer constraints, we propose the KMC+ heuristic. The method builds on classical k-means clustering and appends a post-clustering verification and dimensioning phase. This extension ensures that when optical power splitting ratios are determined, the locations of both the DU and splitter are fixed, enabling accurate evaluation of end-to-end optical path loss and fronthaul latency under physical constraints.

The KMC+ approach relies on a Lloyd-style k-means routine with k-means++ initialization (Algorithm~\ref{alg:kmeans_sub}) to partition points in $\mathbb{R}^2$ by distance minimization. Distances are computed under a Euclidean distance metric.

\begin{algorithm}[t]
\caption{\textsc{KMeans}$(\mathcal{P},K,I_{\mathrm{km}},\varepsilon_{\mathrm{km}};\delta)$ (Lloyd + k-means++)}\label{alg:kmeans_sub}
\begin{algorithmic}[1]
\Require point set $\mathcal{P}=\{p_1,\dots,p_n\}\subset\mathbb{R}^2$, cluster count $K$, max iterations $I_{\mathrm{km}}$, tolerance $\varepsilon_{\mathrm{km}}$, distance metric $\delta(\cdot,\cdot)$
\Ensure centers $\{c_1,\dots,c_K\}$ and assignments $\pi(i)\in\{1,\dots,K\}$
\State Initialize centers $c_1,\dots,c_K$ using k-means++ seeding
\For{$it=1$ to $I_{\mathrm{km}}$}
    \State $\pi(i)\leftarrow \arg\min_{k\in\{1,\dots,K\}} \delta(p_i,c_k) \quad \forall i$
    \State Update centers $c_k \leftarrow \frac{1}{|\{i:\pi(i)=k\}|}\sum_{i:\pi(i)=k} p_i \quad \forall k$
    \State \textbf{(If any cluster is empty)} re-seed its center to a random point in $\mathcal{P}$
    \State $\Delta \leftarrow \max_k \|c_k-c_k^{\mathrm{old}}\|_2$
    \If{$\Delta \le \varepsilon_{\mathrm{km}}$}
        \State \textbf{break}
    \EndIf
\EndFor
\State \Return $\{c_k\}_{k=1}^K$, $\pi(\cdot)$
\end{algorithmic}
\end{algorithm}

\begin{algorithm}[!t]
\caption{Improved K-Means Clustering (KMC+)}\label{alg:KMC+_main_short}
\begin{algorithmic}[1]
\Require $\mathcal{R},\mathcal{S},\mathcal{D}$; physical params $(P^{budget},T^{FH},L^{fib},L^{fix},\Delta L^m,\mathcal{T},v^f,T^{proc},N_{RU}^{max})$;
k-means params $(I_{\mathrm{km}},\varepsilon_{\mathrm{km}})$; controls $(T_{\text{run}},I_{\text{pat}},\epsilon,I_{\text{in}})$; metric $\delta(\cdot,\cdot)$
\Ensure Best feasible assignment $\{(d,s,r,t^*(s))\}$ within $T_{\text{run}}$ (or infeasible)

\State Initialize $K_S \!\leftarrow\! \left\lceil |\mathcal{R}| / \operatorname{avg}_{t\in\mathcal{T}}(2^t)\right\rceil$, \;
$K_D \!\leftarrow\! \left\lceil |\mathcal{R}| /(N_{RU}^{max}/2)\right\rceil$
\State $\text{bestCost}\!\leftarrow\!+\infty$; $c_{\text{noImp}}\!\leftarrow\!0$
\State $t_0 \leftarrow$ current time
\While{$\text{current time} - t_0 <T_{\text{run}}$}
  \For{$\text{in}=1$ to $I_{\text{in}}$} \Comment{retries at fixed $(K_S,K_D)$}
    \State \textbf{Phase 1 (RU$\to$S):} cluster $\mathcal{R}$ into $K_S$ via \textsc{KMeans}; map centroids $\to$ nearest $s\in\mathcal{S}$ to get $s(r)$ and $\mathcal{S}_a$
    \If{$|\mathcal{S}_a|=0$} \State $\text{fail}\!\leftarrow\!\texttt{KS}$; \textbf{continue} \EndIf
    \State \textbf{Phase 2 (S$\to$D):} build weighted set $\widetilde{\mathcal{S}}_a$ by capped replication using RU counts; cluster into $K_D$; map centroids $\to$ nearest $d\in\mathcal{D}$ to get $d(s)$
    \If{no active DU} \State $\text{fail}\!\leftarrow\!\texttt{KD}$; \textbf{continue} \EndIf

    \State \textbf{Phase 3 (check+dimension):} for each $s\in\mathcal{S}_a$ compute
    $d_{\text{total}}(s)=\delta(d(s),s)+\max_{r:s(r)=s}\delta(s,r)$
    \If{$\exists s:\; \frac{1000\,d_{\text{total}}(s)}{v^f}+T^{proc}>T^{FH}$}
        \State $\text{fail}\!\leftarrow\!\texttt{KS}$; \textbf{continue}
    \EndIf
    \If{$\exists s$ with no $t\in\mathcal{T}$ s.t. $L^{fib}d_{\text{total}}(s)+L^{split}_{t}+L^{fix}+\Delta L^m\le P^{budget}$}
        \State $\text{fail}\!\leftarrow\!\texttt{KD}$; \textbf{continue}
    \EndIf
    \State For each $s$: $t^*(s)\leftarrow \min\{\lceil\log_2(N_s)\rceil,\; t_{\mathrm{feas}}(s)\}$,\;
$n_{dst}(s)\leftarrow \left\lceil N_s/2^{t^*(s)}\right\rceil$
    \If{$\exists d:\; |\{r:d(s(r))=d\}|>N_{RU}^{max}$}
        \State $\text{fail}\!\leftarrow\!\texttt{KD}$; \textbf{continue}
    \EndIf
    \If{\textsc{FinalFeasibility} fails} \State $\text{fail}\!\leftarrow\!\texttt{KD}$; \textbf{continue} \EndIf

    \State Compute cost $C$; update best if $C<\text{bestCost}-\epsilon$ else $c_{\text{noImp}}{+}{+}$
    \If{$c_{\text{noImp}}\ge I_{\text{pat}}$} \State \Return best solution \EndIf
  \EndFor
  \State \textbf{Bump rule:} if $\text{fail}=\texttt{KS}$ then $K_S{+}{+}$ else $K_D{+}{+}$ (swap if maxed); stop if both maxed
\EndWhile
\State \Return best solution
\end{algorithmic}
\end{algorithm}

\subsection{Methodology}

The KMC+ heuristic (Algorithm~\ref{alg:KMC+_main_short}) runs until either (i) the time budget $T_{\text{run}}$ is exceeded, or (ii) early stopping triggers after $I_{\text{pat}}$ consecutive feasible solutions that do not improve the best cost by at least $\epsilon$.

\begin{itemize}
    \item \textbf{Phase 1: Splitter siting (RU clustering).}
    We partition the set of RUs $\mathcal{R}$ into $K_S$ clusters using \textsc{KMeans}. The initial value $K_S$ is set via an average splitter fan-out proxy, $\operatorname{avg}_{t\in\mathcal{T}}(2^t)$. Cluster centroids are mapped to the nearest candidate splitter sites $s\in\mathcal{S}$, which induces RU-to-splitter assignments $s(r)$ and the active splitter set $\mathcal{S}_a$.

    \item \textbf{Phase 2: DU siting (splitter clustering with capped replication).}
    Each active splitter $s\in\mathcal{S}_a$ is treated as a demand point weighted by its RU count $N_s$. In implementation, this weight is realized by capped replication of each $s$ to form a weighted set $\widetilde{\mathcal{S}}_a$. Running \textsc{KMeans} with $K_D$ clusters on $\widetilde{\mathcal{S}}_a$ yields centroids mapped to candidate DU sites $d\in\mathcal{D}$, selecting the active DU set $\mathcal{D}_a$. Each active splitter is then assigned to its nearest selected DU $d(s)\in\mathcal{D}_a$.

    \item \textbf{Phase 3: Path verification and dynamic dimensioning.}
    For each active splitter $s$ assigned to a DU $d(s)$ serving RU subset $\mathcal{R}_s=\{r\in\mathcal{R}:s(r)=s\}$, define the worst-case feeder+distribution length
    \begin{equation}
        d_{\text{total}}(s)= \delta(d(s),s) + \max_{r\in\mathcal{R}_s} \delta(s,r).
    \end{equation}
    The algorithm verifies:
    \begin{enumerate}
        \item \textbf{Latency constraint:}
        \begin{equation}
            \frac{1000\, d_{\text{total}}(s)}{v^f} + T^{proc} \le T^{FH}.
        \end{equation}
        If violated for any $s$, the current attempt is rejected and the heuristic prefers increasing $K_S$ (more splitters, smaller clusters).

        \item \textbf{Power budget check (largest feasible splitter type):}
        For splitter types $t\in\mathcal{T}$ (checked in descending capacity order), we verify
        \begin{equation}
            L^{fib}\cdot d_{\text{total}}(s) + L^{split}_{t} + L^{fix} + \Delta L^m \le P^{budget}.
        \end{equation}
        Let $t_{\mathrm{feas}}(s)$ denote the largest $t$ satisfying the inequality. If no $t$ is feasible for some $s$, the attempt is rejected and the heuristic prefers increasing $K_D$ (more DUs, reduced feeder lengths).

        \item \textbf{Dimensioning:}
        Let $N_s = |\mathcal{R}_s|$ and define the minimum required type
        \begin{equation}
            t_{\mathrm{need}}(s) = \left\lceil \log_2(N_s)\right\rceil,\qquad t_{\mathrm{need}}(s)\ge 1.
        \end{equation}
        We select
        \begin{equation}
            t^*(s)=\min\{t_{\mathrm{need}}(s),\, t_{\mathrm{feas}}(s)\},\qquad t^*(s)\ge 1,
        \end{equation}
        and the number of splitters deployed at $s$ is
        \begin{equation}
            n_{dst}(s) = \left\lceil \frac{N_s}{2^{t^*(s)}} \right\rceil.
        \end{equation}
    \end{enumerate}

    \item \textbf{DU capacity constraint:}
    Finally, we verify that the total number of RUs assigned to any DU does not exceed $N_{RU}^{max}$. If violated, the attempt is rejected and the heuristic prefers increasing $K_D$.

    \item \textbf{Final feasibility and selection:}
    After passing the above checks, a full feasibility checker is executed on the constructed tuples $\{(d,s,r,t^*(s))\}$. Feasible solutions are evaluated by total cost $C$, and the best feasible solution is retained. Early stopping triggers when no improvement larger than $\epsilon$ is observed for $I_{\text{pat}}$ consecutive feasible evaluations.
\end{itemize}

\section{Improved Randomized Successive Splitter Assignment (RSSA+)}
\label{sec:graph_approach}

This section presents a constructive heuristic, RSSA+, that builds a feasible fronthaul topology by assigning RUs in a randomized order and greedily selecting, for each RU, the DU--splitter pair $(d,s)$ that minimizes an incremental-cost proxy under the current partial network state. The method explicitly models step-wise shared infrastructure costs (e.g., trenching paid once per feeder corridor and additional splitter instances opened only when needed) and incorporates DU power-level upgrades through a discrete capacity ladder.

RSSA is executed over multiple randomized runs and returns the minimum TCO among feasible runs. Within each run, RSSA uses {randomized category weights} to diversify greedy decisions, but {the final reported objective is always the true unweighted TCO} computed after post-dimensioning.

\subsection{Graph Construction}
\label{app:graph}

We define a directed graph $G=(V,E)$ with four disjoint vertex sets:
\begin{itemize}
    \item $v_{\mathrm{root}}$: a virtual root node representing the network core.
    \item $V_D$: candidate DU pool locations, corresponding to $\mathcal{D}$.
    \item $V_S$: candidate splitter locations, corresponding to $\mathcal{S}$.
    \item $V_R$: RUs, corresponding to $\mathcal{R}$.
\end{itemize}

Edges encode the 3-stage path $v_{\mathrm{root}} \to d \to s \to r$. A candidate connection for RU $r$ is represented by a triple $(d,s,t)$, where $t \in \mathcal{T}=\{1,\dots,T_{\max}\}$ denotes the splitter type with splitting ratio $2^t$. In the implementation used here, greedy assignments are made in a {Phase-1} mode that forces a fixed splitter type $t=t_{\mathrm{ph1}}$ during assignment; after all RUs are assigned, splitters are {post-dimensioned} to a single type per used $(d,s)$.

\subsection*{Feasibility of $v_{\mathrm{root}} \to d \to s \to r$}

A candidate path via $(d,s,t)$ is feasible for RU $r$ only if it satisfies:

\begin{enumerate}
    \item \textbf{Latency:}
    \begin{equation}
        \frac{\delta_{ds} + \delta_{sr}}{v^{\mathrm{f}}} + T^{\mathrm{proc}}_r \le T^{\mathrm{FH}}.
    \end{equation}

    \item \textbf{Optical power budget:}
    \begin{equation}
        L^{\mathrm{fib}}\!\left(\delta_{ds} + \delta_{sr}\right)
        + L^{\mathrm{split}}_{t}
        + L^{\mathrm{fix}}
        + \Delta L^{\mathrm{m}}
        \le P^{\mathrm{budget}}.
    \end{equation}

    \item \textbf{DU capacity feasibility :}
    assigning RU $r$ to DU $d$ must be supportable by at least one DU level in the configured ladder.
\end{enumerate}

\subsection*{Incremental Cost Proxy Used by RSSA (Greedy Selection)}

During a run, RSSA assigns RUs sequentially. For each RU $r$, it evaluates candidate $(d,s)$ pairs under the forced Phase-1 splitter type $t_{\mathrm{ph1}}$ and chooses the minimum value of a {weighted} incremental proxy:
\begin{equation}
    \Delta \widetilde{C}(d,s; r)
    =
    w_1\,\Delta C^{\mathrm{DU}}(d)
    + w_2\,\Delta C^{\mathrm{DS}}(d,s)
    + w_3\,C^{\mathrm{SR}}(s,r),
\end{equation}
where $(w_1,w_2,w_3)$ are sampled i.i.d.\ per run from a uniform random vector and normalized so that $w_1+w_2+w_3=1$.
These weights are used {only} to diversify greedy choices; the final run cost is computed using the true unweighted TCO.

\textbf{(i) RU-splitter cost, $C^{\mathrm{SR}}(s,r)$:}

The per-RU RU-splitter term used in the greedy proxy is:
\begin{equation}
    C^{\mathrm{SR}}(s,r)
    =
    \delta_{sr}\left(C^{\mathrm{tr}} + C^{\mathrm{df}}\right)
    + C^{\mathrm{onu}},
\end{equation}
where $C^{\mathrm{onu}}$ is selected from the ONU-rate table based on the per-RU bitrate requirement.

\textbf{(ii) DU-splitter incremental cost, $\Delta C^{\mathrm{DS}}(d,s)$:}

This term models trenching as a one-time corridor cost and models opening additional splitter {instances} via a step cost when the currently opened instances at $(d,s)$ have no remaining ports.

Let $\mathrm{trenchPaid}_{ds}\in\{0,1\}$ indicate whether trenching for corridor $(d,s)$ has already been paid. The one-time trenching term is:
\begin{equation}
    \Delta C^{\mathrm{DS,trench}}(d,s)
    =
    (1-\mathrm{trenchPaid}_{ds})\,\delta_{ds}\,C^{\mathrm{tr}}.
\end{equation}

In Phase-1, all assignments use $t=t_{\mathrm{ph1}}$ and RSSA maintains the remaining open port capacity at $(d,s)$ for this type. If the currently opened splitter instances have at least one free port, no new instance is opened; otherwise, opening one additional instance incurs:
\begin{equation}
    \Delta C^{\mathrm{DS,open}}(d,s)
    =
    \delta_{ds}\,C^{\mathrm{ff}}
    + C^{\mathrm{sp}}_{t_{\mathrm{ph1}}}\left(1 + T^{\mathrm{op}} C^{\mathrm{m}}\right),
\end{equation}
where $C^{\mathrm{m}}$ is the annual maintenance fraction applied to equipment cost, and $T^{\mathrm{op}}$ is the operating horizon in years.

Thus, the feeder incremental term used by the greedy proxy can be summarized as:
\begin{equation}
    \Delta C^{\mathrm{DS}}(d,s)
    =
    \Delta C^{\mathrm{DS,trench}}(d,s)
    +
    \mathbb{I}[\text{no free port at }(d,s)]\;\Delta C^{\mathrm{DS,open}}(d,s).
\end{equation}

\textbf{(iii) DU incremental cost, $\Delta C^{\mathrm{DU}}(d)$:}

The DU incremental cost has two parts:
\begin{itemize}
    \item a one-time activation cost if $n_d=0$:
    \begin{equation}
        C^{\mathrm{DU,open}}(d)
        =
        C^{\mathrm{bp}}\left(1 + T^{\mathrm{op}} C^{\mathrm{m}}\right)
        + T^{\mathrm{op}} C^{\mathrm{rent}},
    \end{equation}
    \item an incremental energy OpEx if the additional RU forces a higher DU power level.
\end{itemize}

\subsection{Post-dimensioning After Greedy Assignment}

After all RUs are assigned (using $t_{\mathrm{ph1}}$ during assignment), RSSA post-dimensions:
    Let $N_{ds}$ be the number of RUs assigned through $(d,s)$.
    For each RU assignment, the feasibility routine yields a per-RU maximum feasible splitter type at $(d,s)$; RSSA stores the group feasibility bound
    \begin{equation}
        t^{\max}_{ds} = \min_{r \in \mathcal{R}_{ds}} t^{\max}_{ds}(r),
    \end{equation}
    where $\mathcal{R}_{ds}$ is the set of RUs assigned to $(d,s)$ and $t^{\max}_{ds}(r)$ is the maximum feasible $t$ for RU $r$ at $(d,s)$.

    The post-dimensioned splitter type is then chosen demand-aware and feasibility-aware:
    \begin{equation}
        t^{\mathrm{need}}_{ds} = \max\left\{1,\left\lceil \log_2(N_{ds}) \right\rceil\right\},
        \qquad
        t^\star_{ds} = \min\{t^{\max}_{ds},\,t^{\mathrm{need}}_{ds}\},
    \end{equation}
    with $t^\star_{ds}$ clipped to $[1,T_{\max}]$ for used sites.

    The number of splitter instances at $(d,s)$ is:
    \begin{equation}
        n_{ds} = \left\lceil \frac{N_{ds}}{2^{t^\star_{ds}}} \right\rceil.
    \end{equation}
    Finally, all RUs assigned to $(d,s)$ are updated to report splitter type $t^\star_{ds}$.

\begin{algorithm}[!ht]
\caption{Improved Randomized Successive Splitter Assignment (RSSA+)}
\label{alg:rssa}
\begin{algorithmic}[1]
\renewcommand{\algorithmicrequire}{\textbf{Input:}}
\renewcommand{\algorithmicensure}{\textbf{Output:}}
\Require Sets $\mathcal{D},\mathcal{S},\mathcal{R},\mathcal{T}$; distances $\delta_{ds},\delta_{sr}$; parameters and costs; Phase-1 type $t_{\mathrm{ph1}}$; time limit; $\epsilon$; patience $I_{\mathrm{pat}}$
\Ensure Best feasible assignment and minimum post-dimensioned unweighted TCO
\State $Z_{\mathrm{best}} \leftarrow \infty$, $\mathrm{Sol}_{\mathrm{best}} \leftarrow \emptyset$
\State $\mathrm{noImprove}\leftarrow 0$, $\mathrm{runIdx}\leftarrow 0$, start timer
\While{time remaining \textbf{and} $\mathrm{noImprove} < I_{\mathrm{pat}}$}
    \State Sample random weights $w\sim U(0,1)^3$ and normalize so $\sum_i w_i=1$
    \State Initialize assignment arrays (unset), and state:
    \State \quad $\mathrm{trenchPaid}_{ds}\leftarrow 0$; remaining splitter ports $\mathrm{remCap}_{ds}\leftarrow 0$
    \State \quad DU active flags $\mathrm{active}_d\leftarrow 0$; DU loads $n_d\leftarrow 0$; DU levels $k_d\leftarrow 0$
    \State Randomize RU order $\pi \leftarrow \mathrm{RandomPermutation}(\mathcal{R})$
    \State $\mathrm{feasibleRun}\leftarrow \mathrm{true}$
    \For{each $r$ in order $\pi$}
        \State Compute feasibility over $(d,s,t)$ to obtain $t^{\max}_{ds}(r)$ (latency+power); infeasible $\to 0$
        \State Restrict candidates to $(d,s)$ with $t^{\max}_{ds}(r)\ge t_{\mathrm{ph1}}$
        \If{no feasible $(d,s)$ exists}
            \State $\mathrm{feasibleRun}\leftarrow \mathrm{false}$; \textbf{break}
        \EndIf
        \State Compute DU incremental cost $\Delta C^{\mathrm{DU}}(d)$ for adding RU to each $d$ (including level upgrade energy)
        \State Form base weighted score over $(d,s)$:
        \State \quad $B_{ds} \leftarrow w_1\Delta C^{\mathrm{DU}}(d) + w_2\,\Delta C^{\mathrm{DS,trench}}(d,s) + w_3\,C^{\mathrm{SR}}(s,r)$
        \State If $\mathrm{remCap}_{ds}>0$, candidate score is $B_{ds}$; else score is $B_{ds} + w_2\,\Delta C^{\mathrm{DS,open}}(d,s)$
        \State Choose $(d^\star,s^\star) \leftarrow \arg\min$ feasible candidate score; set $t^\star \leftarrow t_{\mathrm{ph1}}$
        \State Commit assignment of RU $r$ to $(d^\star,s^\star,t^\star)$
        \State Update corridor state: if $\mathrm{trenchPaid}_{d^\star s^\star}=0$, set it to $1$
        \State Update splitter capacity: if no free port, open one instance; then consume one port
        \State Update group feasibility bound at $(d^\star,s^\star)$: $t^{\max}_{d^\star s^\star} \leftarrow \min(\cdot)$
        \State Update DU load $n_{d^\star}\leftarrow n_{d^\star}+1$ and DU level $k_{d^\star}$ (validate capacity)
        \If{DU capacity violated}
            \State $\mathrm{feasibleRun}\leftarrow \mathrm{false}$; \textbf{break}
        \EndIf
    \EndFor
    \If{\textbf{not} $\mathrm{feasibleRun}$}
        \State $\mathrm{runIdx}\leftarrow \mathrm{runIdx}+1$; $\mathrm{noImprove}\leftarrow \mathrm{noImprove}+1$
        \State \textbf{continue}
    \EndIf

    \State \textbf{Post-dimensioning:} for each used $(d,s)$ compute $t^\star_{ds}$ and $n_{ds}$ from load $N_{ds}$ and group bound $t^{\max}_{ds}$
    \State Set each RU on $(d,s)$ to report $t^\star_{ds}$; compute DU levels $k^\star_d$ from final DU loads
    \State Compute \textbf{unweighted} post-dimensioned TCO $Z$ (true objective)
    \If{$Z < Z_{\mathrm{best}} - \epsilon$}
        \State $Z_{\mathrm{best}}\leftarrow Z$; store solution; $\mathrm{noImprove}\leftarrow 0$
    \Else
        \State $\mathrm{noImprove}\leftarrow \mathrm{noImprove}+1$
    \EndIf
    \State $\mathrm{runIdx}\leftarrow \mathrm{runIdx}+1$
\EndWhile
\State \Return stored best feasible solution and $Z_{\mathrm{best}}$
\end{algorithmic}
\end{algorithm}

%% file: references.bib
@article{6GVision,
  author    = {Dang, S. and Amin, O. and Shihada, B. and Alouini, M.-S.},
  title     = {What Should {6G} Be?},
  journal   = {Nature Electronics},
  volume    = {3},
  number    = {1},
  pages     = {20--29},
  year      = {2020}
}

@ARTICLE{10546919,
  author    = {Fayad, Abdulhalim and Cinkler, Tibor and Rak, Jacek},
  journal   = {IEEE Communications Surveys \& Tutorials}, 
  title     = {Toward {6G} Optical Fronthaul: A Survey on Enabling Technologies and Research Perspectives}, 
  year      = {2025},
  volume    = {27},
  number    = {1},
  pages     = {629-666},
  doi       = {10.1109/COMST.2024.3408090}
}

@article{Checko2015,
  author    = {Checko, Aleksandra and Christiansen, Henrik L. and Yan, Ying and Scolari, Lara and Kardaras, Georgios and Berger, Michael S. and Dittmann, Lars},
  journal   = {IEEE Communications Surveys \& Tutorials}, 
  title     = {{Cloud RAN} for Mobile Networks—A Technology Overview}, 
  year      = {2015},
  volume    = {17},
  number    = {1},
  pages     = {405-426},
  doi       = {10.1109/COMST.2014.2355255}
}

@article{maes2024efficient,
  title     = {Efficient transport of enhanced {CPRI} fronthaul over {PON}},
  author    = {Maes, Jochen and Bidkar, Sarvesh and Straub, Michael and Pfeiffer, Thomas and Bonk, Rene},
  journal   = {Journal of Optical Communications and Networking},
  volume    = {16},
  number    = {2},
  pages     = {A136--A142},
  year      = {2024},
  publisher = {Optica Publishing Group}
}

@techreport{hughes2021planning,
  title={Planning and design considerations for data centers},
  author={Hughes, Lyndsi and Sweeney, David and Kasunic, Mark},
  year={2021},
  institution={Technical note CMU/SEI-2021-TN-002}
}

@techreport{3GPP-38801,
  author      = {3GPP},
  title       = {Study on New Radio Access Technology: Radio Access Architecture and Interfaces},
  institution = {3rd Generation Partnership Project ({3GPP})},
  number      = {TR 38.801 V14.0.0},
  year        = {2017}
}

@misc{ORAN-WG4,
  title        = {{O-RAN} Fronthaul Working Group 4: Control, User and Synchronization Plane Specification v7.00},
  author       = {{O-RAN Alliance}},
  year         = {2022},
  howpublished = {\url{https://www.o-ran.org/}},
  note         = {Accessed: 2025-12-13}
}

@ARTICLE{11006330,
  author    = {Dass, Devika and Kilper, Dan and Barry, Liam and Ruffini, Marco},
  journal   = {Journal of Optical Communications and Networking}, 
  title     = {Heterogeneous transmission of analog radio and digital coherent signals over multispan metro and {PON} for bandwidth-efficient fronthaul in mmWave centralized {RAN}s [Invited]}, 
  year      = {2025},
  volume    = {17},
  number    = {8},
  pages     = {C136-C143},
  doi       = {10.1364/JOCN.551296}
}

@ARTICLE{11172732,
  author    = {Vujicic, Zoran and Gelabert, Xavier and Santos, Maria C. and Mendez, Rodrigo and Gaudino, Roberto},
  journal   = {Journal of Optical Communications and Networking}, 
  title     = {On the feasibility of analog fiber dispersion-based photonic beamforming for mmWave wireless access over 100{GHz} {DWDM} grids}, 
  year      = {2025},
  volume    = {17},
  number    = {11},
  pages     = {E70-E81},
  doi       = {10.1364/JOCN.567208}
}

@article{horvath2020passive,
  title     = {Passive optical networks progress: a tutorial},
  author    = {Horvath, Tomas and Munster, Petr and Oujezsky, Vaclav and Bao, Ning-Hai},
  journal   = {Electronics},
  volume    = {9},
  number    = {7},
  pages     = {1081},
  year      = {2020},
  publisher = {MDPI}
}

@ARTICLE{11021422,
  author    = {Akhtar, Md Shahbaz and Kumar, Mohit and Iftekhar Alam, Md and Adhya, Aneek},
  journal   = {IEEE Transactions on Network and Service Management}, 
  title     = {{XGS-PON}-Standard Compliant {DBA} Algorithm for Option 7.x Functional Split-Based {5G} {C-RAN}}, 
  year      = {2025},
  volume    = {22},
  number    = {5},
  pages     = {5048-5061},
  doi       = {10.1109/TNSM.2025.3575938}
}

@article{Nakayama2019,
  title     = {Wavelength and bandwidth allocation for mobile fronthaul in {TWDM-PON}},
  author    = {Nakayama, Yu and Hisano, Daisuke},
  journal   = {IEEE Transactions on Communications},
  volume    = {67},
  number    = {11},
  pages     = {7642--7655},
  year      = {2019},
  publisher = {IEEE}
}

@article{Chen2016JOCN,
  author    = {Chen, Hao and Li, Yongcheng and Bose, Sanjay K. and Shao, Weidong and Xiang, Lian and Ma, Yiran and Shen, Gangxiang},
  journal   = {Journal of Optical Communications and Networking}, 
  title     = {Cost-minimized design for {TWDM-PON}-based {5G} mobile backhaul networks}, 
  year      = {2016},
  volume    = {8},
  number    = {11},
  pages     = {B1-B11},
  doi       = {10.1364/JOCN.8.0000B1}
}

@article{ranaweera20175g,
  title     = {{5G} {C-RAN} with optical fronthaul: An analysis from a deployment perspective},
  author    = {Ranaweera, Chathurika and Wong, Elaine and Nirmalathas, Ampalavanapillai and Jayasundara, Chamil and Lim, Christina},
  journal   = {Journal of Lightwave Technology},
  volume    = {36},
  number    = {11},
  pages     = {2059--2068},
  year      = {2017},
  publisher = {IEEE}
}

@article{ranaweera2019optical,
  title     = {Optical transport network design for {5G} fixed wireless access},
  author    = {Ranaweera, Chathurika and Monti, Paolo and Skubic, Bj{\"o}rn and Wong, Elaine and Furdek, Marija and Wosinska, Lena and Machuca, Carmen Mas and Nirmalathas, Ampalavanapillai and Lim, Christina},
  journal   = {Journal of Lightwave Technology},
  volume    = {37},
  number    = {16},
  pages     = {3893--3901},
  year      = {2019},
  publisher = {OSA}
}

@article{musumeci2016optimal,
  title     = {Optimal {BBU} placement for {5G} {C-RAN} deployment over {WDM} aggregation networks},
  author    = {Musumeci, Francesco and Bellanzon, Camilla and Carapellese, Nicola and Tornatore, Massimo and Pattavina, Achille and Gosselin, St{\'e}phane},
  journal   = {Journal of Lightwave Technology},
  volume    = {34},
  number    = {8},
  pages     = {1963--1970},
  year      = {2016},
  publisher = {OSA}
}

@article{dias2022evolutionary,
  title     = {Evolutionary strategy for practical design of passive optical networks},
  author    = {Dias, Leonardo Pereira and Dos Santos, Alex Ferreira and Pereira, Helder Alves and de Andrade Almeida Jr, Raul Camelo and Giozza, William Ferreira and de Sousa Jr, Rafael Tim{\'o}teo and Assis, Karcius Day Rosario},
  journal = {Photonics},
  volume    = {9},
  number    = {5},
  pages     = {278},
  year      = {2022},
  organization = {MDPI}
}

@article{akhtar2023fronthaul,
  title     = {Fronthaul latency and capacity constrained cost-effective and energy-efficient {5G} {C-RAN} deployment},
  author    = {Akhtar, Md Shahbaz and Gupta, Jitendra and Alam, Md Iftekhar and Majhi, Sudhan and Adhya, Aneek},
  journal   = {Optical Fiber Technology},
  volume    = {80},
  pages     = {103392},
  year      = {2023},
  publisher = {Elsevier}
}

@inproceedings{lisi2017cost,
  title     = {Cost-effective migration towards {C-RAN} with optimal fronthaul design},
  author    = {Lisi, Shari Sofia and Alabbasi, Abdulrahman and Tornatore, Massimo and Cavdar, Cicek},
  booktitle = {2017 IEEE International Conference on Communications (ICC)},
  pages     = {1--7},
  year      = {2017},
  organization = {IEEE}
}

@article{masoudi2020cost,
  title     = {Cost-effective migration toward virtualized {C-RAN} with scalable fronthaul design},
  author    = {Masoudi, Meysam and Lisi, Shari Sofia and Cavdar, Cicek},
  journal   = {IEEE Systems Journal},
  volume    = {14},
  number    = {4},
  pages     = {5100--5110},
  year      = {2020},
  publisher = {IEEE}
}

@inproceedings{carapellese2015bbu,
  title     = {{BBU} placement over a {WDM} aggregation network considering {OTN} and overlay fronthaul transport},
  author    = {Carapellese, Nicola and Tornatore, Massimo and Pattavina, Achille and Gosselin, St{\'e}phane},
  booktitle = {2015 European Conference on Optical Communication (ECOC)},
  pages     = {1--3},
  year      = {2015},
  organization = {IEEE}
}

@article{kokangul2011optimization,
  title     = {Optimization of passive optical network planning},
  author    = {Kokangul, A and Ari, A},
  journal   = {Applied Mathematical Modelling},
  volume    = {35},
  number    = {7},
  pages     = {3345--3354},
  year      = {2011},
  publisher = {Elsevier}
}

@article{fayad2022design,
  title     = {Design of cost-efficient optical fronthaul for {5G}/{6G} networks: An optimization perspective},
  author    = {Fayad, Abdulhalim and Cinkler, Tibor and Rak, Jacek and Jha, Manish},
  journal   = {Sensors},
  volume    = {22},
  number    = {23},
  pages     = {9394},
  year      = {2022},
  publisher = {MDPI}
}

@article{fayad20235g,
  title     = {{5G}/{6G} optical fronthaul modeling: Cost and energy consumption assessment},
  author    = {Fayad, Abdulhalim and Cinkler, Tibor and Rak, Jacek},
  journal   = {Journal of Optical Communications and Networking},
  volume    = {15},
  number    = {9},
  pages     = {D33--D46},
  year      = {2023},
  publisher = {Optica Publishing Group}
}

@manual{cpri_specification_70,
  author       = {{CPRI Cooperation}},
  title        = {Common Public Radio Interface ({CPRI}) Specification v7.0},
  institution  = {CPRI Cooperation},
  type         = {Standard},
  number       = {CPRI v7.0},
  year         = {2014},
  url          = {https://www.cpri.info/spec.html}
}

@manual{ecpri_specification_20,
  author       = {{CPRI Cooperation}},
  title        = {{eCPRI} Specification v2.0},
  institution  = {CPRI Cooperation},
  type         = {Standard},
  number       = {eCPRI v2.0},
  year         = {2019},
  url          = {https://www.cpri.info/downloads/eCPRI_v_2.0_2019_05_10c.pdf}
}

@article{wypior2022open,
  title     = {{Open RAN}—radio access network evolution, benefits and market trends},
  author    = {Wypi{\'o}r, Dariusz and Klinkowski, Miros{\l}aw and Michalski, Igor},
  journal   = {Applied Sciences},
  volume    = {12},
  number    = {1},
  pages     = {408},
  year      = {2022},
  publisher = {MDPI}
}

@article{habibi2019comprehensive,
  title     = {A comprehensive survey of {RAN} architectures toward {5G} mobile communication system},
  author    = {Habibi, Mohammad Asif and Nasimi, Meysam and Han, Bin and Schotten, Hans D},
  journal   = {Ieee Access},
  volume    = {7},
  pages     = {70371--70421},
  year      = {2019},
  publisher = {IEEE}
}

@article{polese2023understanding,
  title     = {Understanding {O-RAN}: Architecture, interfaces, algorithms, security, and research challenges},
  author    = {Polese, Michele and Bonati, Leonardo and D’oro, Salvatore and Basagni, Stefano and Melodia, Tommaso},
  journal   = {IEEE Communications Surveys \& Tutorials},
  volume    = {25},
  number    = {2},
  pages     = {1376--1411},
  year      = {2023},
  publisher = {IEEE}
}

@article{bonati2021intelligence,
  title     = {Intelligence and learning in {O-RAN} for data-driven {NextG} cellular networks},
  author    = {Bonati, Leonardo and D'Oro, Salvatore and Polese, Michele and Basagni, Stefano and Melodia, Tommaso},
  journal   = {IEEE Communications Magazine},
  volume    = {59},
  number    = {10},
  pages     = {21--27},
  year      = {2021},
  publisher = {IEEE}
}

@article{liyanage2023open,
  title     = {{Open RAN} security: Challenges and opportunities},
  author    = {Liyanage, Madhusanka and Braeken, An and Shahabuddin, Shahriar and Ranaweera, Pasika},
  journal   = {Journal of Network and Computer Applications},
  volume    = {214},
  pages     = {103621},
  year      = {2023},
  publisher = {Elsevier}
}

@article{larsen2018survey,
  title     = {A survey of the functional splits proposed for {5G} mobile crosshaul networks},
  author    = {Larsen, Line MP and Checko, Aleksandra and Christiansen, Henrik L},
  journal   = {IEEE Communications Surveys \& Tutorials},
  volume    = {21},
  number    = {1},
  pages     = {146--172},
  year      = {2018},
  publisher = {IEEE}
}

@article{dao2024review,
  title     = {A review on new technologies in {3GPP} standards for {5G} access and beyond},
  author    = {Dao, Nhu-Ngoc and Tu, Ngo Hoang and Hoang, Trong-Dai and Nguyen, Tri-Hai and Nguyen, Luong Vuong and Lee, Kyungchun and Park, Laihyuk and Na, Woongsoo and Cho, Sungrae},
  journal   = {Computer Networks},
  volume    = {245},
  pages     = {110370},
  year      = {2024},
  publisher = {Elsevier}
}

@article{liu2022enabling,
  title     = {Enabling optical network technologies for {5G} and beyond},
  author    = {Liu, Xiang},
  journal   = {Journal of Lightwave Technology},
  volume    = {40},
  number    = {2},
  pages     = {358--367},
  year      = {2022},
  publisher = {OSA}
}

@misc{feng2023key,
  title     = {Key Technologies for a {Beyond-100G} Next-Generation Passive Optical Network. Photonics 2023, 10, 1128},
  author    = {Feng, N and Ma, M and Zhang, Y and Tan, X and Li, Z and Li, S},
  year      = {2023}
}

@inproceedings{erbayat2024fronthaul,
  title     = {Fronthaul Network Architecture and Design For Optically Powered Passive Optical Networks},
  author    = {Erbayat, Egemen and Petale, Shrinivas and Lin, Shih-Chun and Matsuura, Motoharu and Hasegawa, Hiroshi and Subramaniam, Suresh},
  booktitle = {ICC 2024-IEEE International Conference on Communications},
  pages     = {4961--4966},
  year      = {2024},
  organization = {IEEE}
}

@inproceedings{erbayat2024design,
  title     = {Design Of Passive Optically-Powered Fronthaul Networks: The {Multi-Splitter} Case},
  author    = {Erbayat, Egemen and Figueiredo, Gustavo B and Lin, Shih-Chun and Matsuura, Motoharu and Hasegawa, Hiroshi and Subramaniam, Suresh},
  booktitle = {2024 IEEE Future Networks World Forum (FNWF)},
  pages     = {495--500},
  year      = {2024},
  organization = {IEEE}
}

@article{erbayat2025toward,
  title     = {Toward scalable passive optically powered fronthaul networks},
  author    = {Erbayat, Egemen and Figueiredo, Gustavo B and Petale, Shrinivas and Lin, Shih-Chun and Matsuura, Motoharu and Hasegawa, Hiroshi and Subramaniam, Suresh},
  journal   = {Journal of Optical Communications and Networking},
  volume    = {17},
  number    = {9},
  pages     = {D125--D136},
  year      = {2025},
  publisher = {Optica Publishing Group}
}

@inproceedings{erbayat2025multi,
  title     = {Multi-{DU} fronthaul design leveraging power over fiber},
  author    = {Erbayat, Egemen and Figueiredo, Gustavo B and Lin, Shih-Chun and Matsuura, Motoharu and Hasegawa, Hiroshi and Subramaniam, Suresh},
  booktitle = {2025 International Conference on Optical Network Design and Modeling (ONDM)},
  year      = {2025}
}

@inproceedings{erbayat2025latency,
  title     = {Latency-Aware Fronthaul Network Design Using {Power-over-Fiber}},
  author    = {Erbayat, Egemen and Dos Santos, Luiz GS and Figueiredo, Gustavo B and Lin, Shih-Chun and Matsuura, Motoharu and Hasegawa, Hiroshi and Subramaniam, Suresh},
  booktitle = {2025 25th Anniversary International Conference on Transparent Optical Networks (ICTON)},
  pages     = {1--4},
  year      = {2025},
  organization = {IEEE}
}

@inproceedings{erbayat2025design,
  title     = {Design of Survivable Fronthaul Networks with {Power-over-Fiber}},
  author    = {Erbayat, Egemen and Figueiredo, Gustavo B and Lin, Shih-Chun and Matsuura, Motoharu and Hasegawa, Hiroshi and Subramaniam, Suresh},
  booktitle = {2025 IEEE Future Networks World Forum (FNWF)},
  pages     = {},
  year      = {2025},
  organization = {IEEE}
}

@article{tinini2020energy,
  title     = {Energy-efficient {vBBU} migration and wavelength reassignment in {Cloud-Fog RAN}},
  author    = {Tinini, Rodrigo Izidoro and Batista, Daniel Mac{\^e}do and Figueiredo, Gustavo Bittencourt and Tornatore, Massimo and Mukherjee, Biswanath},
  journal   = {IEEE Transactions on Green Communications and Networking},
  volume    = {5},
  number    = {1},
  pages     = {18--28},
  year      = {2020},
  publisher = {IEEE}
}

@article{ciceri2024resource,
  title     = {Resource allocation in passive optical networks for low-latency mobile fronthauling services},
  author    = {Ciceri, Oscar J and Astudillo, Carlos A and Figueiredo, Gustavo B and Zhu, Zuqing and Da Fonseca, Nelson LS},
  journal   = {IEEE Network},
  year      = {2024},
  publisher = {IEEE}
}

@article{de2024radio,
  title={Radio-and Power-over-Fiber Integration for {6G} Networks: Challenges and Future Prospects},
  author={De Souza, Let{\'\i}cia Carneiro and Souto, Victoria Dala Pegorara and Sodr{\'e}, Arismar Cerqueira},
  journal={{IEEE} Access},
  year={2024},
  publisher={IEEE}
}

@article{liu2023machine,
  title={Machine Learning for {6G} Enhanced Ultra-Reliable and Low-Latency Services},
  author={Liu, Yan and Deng, Yansha and Nallanathan, Arumugam and Yuan, Jinhong},
  journal={{IEEE} Wireless Communications},
  volume={30},
  number={2},
  pages={48--54},
  year={2023},
  publisher={IEEE}
}

@inproceedings{erbayat2024trade,
  title={A Trade-off Analysis of Latency, Accuracy, and Energy in Task Offloading Strategies for {UAVs}},
  author={Erbayat, Egemen and Zou, Rujia and Wei, Xianglin and Venkataramani, Guru and Subramaniam, Suresh},
  booktitle={2024 IEEE Cloud Summit},
  pages={48--53},
  year={2024},
  organization={IEEE}
}

@misc{technicalNote,
  author       = {Erbayat, Egemen and Figueiredo, Gustavo B. and Lin, Shih-Chun and Matsuura, Motoharu and Hasegawa, Hiroshi and Subramaniam, Suresh},
  title        = {{PON-FD}: Benchmarking Framework for {PON}-Based Fronthaul Network Design},
  year         = {2025},
  howpublished = {\url{https://github.com/erbayat/PON-FD-Fronthaul-Benchmarking-Framework}},
  note         = {Open-source benchmarking framework},
}

@techreport{FBA_Cartesian_2025,
  title        = {Fiber Deployment Cost Annual Report 2024},
  author       = {{Fiber Broadband Association} and Cartesian},
  institution  = {Fiber Broadband Association},
  year         = {2025},
  url          = {https://fiberbroadband.org/wp-content/uploads/2025/01/FBA_Cartesian_Fiber-Deployment-Cost-Annual-Report-2024.pdf},
  note         = {Accessed 2025},
  type         = {Annual Industry Cost Study},
}

@article{arevalo2017optimization,
  title={Optimization of Multiple {PON} Deployment Costs and Comparison Between {GPON}, {XGPON}, {NGPON2} and {UDWDM} {PON}},
  author={Ar{\'e}valo, Germ{\'a}n V and Hincapi{\'e}, Roberto C and Gaudino, Roberto},
  journal={Optical Switching and Networking},
  volume={25},
  pages={80--90},
  year={2017},
  publisher={Elsevier}
}

@misc{iea_energy_prices,
  author       = {{International Energy Agency}},
  title        = {Energy Prices},
  year         = {2025},
  howpublished = {\url{https://www.iea.org/data-and-statistics/data-product/energy-prices}},
  note         = {Accessed: 2025-12-18},
  organization = {International Energy Agency}
}

@techreport{ITU_T_G_671,
  author      = {{International Telecommunication Union}},
  title       = {{ITU-T} Recommendation {G.671}: Transmission Characteristics of Optical Components and Subsystems},
  institution = {International Telecommunication Union},
  type        = {Recommendation},
  number      = {ITU-T G.671},
  year        = {2019},
  month       = {August},
  url         = {https://www.itu.int/rec/T-REC-G.671}
}

@techreport{ITU_T_G9807,
  author      = {{International Telecommunication Union}},
  title       = {{ITU-T} Recommendation {G.9807.1}: 10-Gigabit-Capable Symmetric Passive Optical Network ({XGS-PON})},
  institution = {International Telecommunication Union},
  type        = {Recommendation},
  number      = {ITU-T G.9807.1},
  year        = {2023},
  url         = {https://www.itu.int/rec/T-REC-G.9807.1-202505-I!Amd1/},
  note        = {Amendment 1 (2025) in force}
}

@techreport{ITU_T_G987_2_2023_amd1,
  author      = {{International Telecommunication Union}},
  title       = {{ITU-T} Recommendation {G.987.2}: 10-Gigabit-Capable Passive Optical Networks ({XG-PON}): Physical Media Dependent ({PMD}) Layer Specification},
  institution = {International Telecommunication Union},
  type        = {Recommendation},
  number      = {ITU-T G.987.2},
  year        = {2023},
  month       = {June},
  url         = {https://www.itu.int/rec/T-REC-G.987.2-202306-I%21Amd1/},
  note        = {Amendment 1 (2023)}
}

@techreport{ITU_T_G984_2_2019,
  author      = {{International Telecommunication Union}},
  title       = {{ITU-T} Recommendation {G.984.2}: Gigabit-Capable Passive Optical Networks ({GPON}): Physical Media Dependent ({PMD}) Layer Specification},
  institution = {International Telecommunication Union},
  type        = {Recommendation},
  number      = {ITU-T G.984.2},
  year        = {2019},
  month       = {August},
  url         = {https://www.itu.int/rec/T-REC-G.984.2-201908-I/}
}

@techreport{ITU-R-2370,
  author      = {{International Telecommunication Union}},
  title       = {{ITU-R} Recommendation {M.2083-0}: {IMT} Vision -- Framework and Overall Objectives of the Future Development of {IMT} for 2020 and Beyond},
  institution = {International Telecommunication Union},
  type        = {Recommendation},
  number      = {ITU-R M.2083-0},
  year        = {2015},
  month       = {September},
  url         = {https://www.itu.int/rec/R-REC-M.2083/en}
}

@techreport{ITUTG9804,
  author      = {{International Telecommunication Union}},
  title       = {{ITU-T} Recommendation {G.9804.3}: 50-Gigabit-Capable Passive Optical Networks ({50G-PON}): Physical Media Dependent ({PMD}) Layer Specification},
  institution = {International Telecommunication Union},
  type        = {Recommendation},
  number      = {ITU-T G.9804.3},
  year        = {2021},
  address     = {Geneva},
  note        = {Official standard defining 50Gbps symmetric/asymmetric line rates}
}

@techreport{ITUTG989,
  author      = {{International Telecommunication Union}},
  title       = {{ITU-T} Recommendation {G.989.2}: 40-Gigabit-Capable Passive Optical Networks 2 ({NG-PON2}): Physical Media Dependent ({PMD}) Layer Specification},
  institution = {International Telecommunication Union},
  type        = {Recommendation},
  number      = {ITU-T G.989.2},
  year        = {2019},
  address     = {Geneva}
}

@techreport{sup6620205g,
  author      = {{International Telecommunication Union}},
  title       = {{ITU-T} G-series Recommendations -- Supplement 66: {5G} Wireless Fronthaul Requirements in a Passive Optical Network Context},
  institution = {International Telecommunication Union},
  type        = {Supplement},
  number      = {ITU-T G.Sup66},
  year        = {2020},
  month       = {September},
  address     = {Geneva}
}

@techreport{ITURM2412,
  author      = {{International Telecommunication Union}},
  title       = {{ITU-R} Report {M.2412-0}: Guidelines for evaluation of radio interface technologies for {IMT-2020}},
  institution = {International Telecommunication Union},
  type        = {Report}, 
  number      = {ITU-R M.2412-0},
  year        = {2017},
  address     = {Geneva}
}

@techreport{ITURM2160,
  author      = {{International Telecommunication Union}},
  title       = {{ITU-R} Recommendation {M.2160-0}: Framework and overall objectives of the future development of {IMT}-2030 and beyond},
  institution = {International Telecommunication Union},
  type        = {Recommendation}, 
  number      = {ITU-R M.2160-0},
  year        = {2023},}

@techreport{ITURM2516,
  author      = {{International Telecommunication Union}},
  title       = {{ITU-R} Report {M.2516-0}: Future technology trends of terrestrial {International Mobile Telecommunications} systems towards 2030 and beyond},
  institution = {International Telecommunication Union},
  type        = {Report}, 
  number      = {ITU-R M.2516-0},
  year        = {2022},
  month       = {November},
  address     = {Geneva}
}

@misc{arxiv,
  author       = {Egemen Erbayat and Gustavo B. Figueiredo and Shih-Chun Lin and Motoharu Matsuura and Hiroshi Hasegawa and Suresh Subramaniam},
  title        = {A Benchmarking Framework for {PON}-Based Fronthaul Network Design},
  year         = {2026},
  howpublished = {arXiv preprint arXiv:2601.14480},
  note         = {Available: https://arxiv.org/abs/2601.14480}
}

@misc{panitsas20255gcbench,
  author        = {Panitsas, Ioannis and others},
  title         = {{5GC-Bench}: A Framework for Stress-Testing and Benchmarking {5G} Core {VNFs}},
  year          = {2025},
  eprint        = {2509.18443},
  archivePrefix = {arXiv},
  primaryClass  = {cs.NI},
  url           = {https://arxiv.org/abs/2509.18443}
}

@article{natalino2024gym,
  author  = {Natalino, Carlos and Magalh{\~a}es, Talles and Arpanaei, Farhad and Lobato, Fabricio R.~L. and Costa, Jo{\~a}o C.~W.~A. and Hern{\'a}ndez, Jos{\'e} Alberto and Monti, Paolo},
  title   = {{Optical Networking Gym}: an open-source toolkit for resource assignment problems in optical networks},
  journal = {Journal of Optical Communications and Networking},
  volume  = {16},
  number  = {12},
  pages   = {G40--G51},
  year    = {2024},
  doi     = {10.1364/JOCN.532850}
}

@article{pintorios2024benchmarking,
  author  = {Pinto-R{\'\i}os, Juan and Dumas Feris, B{\'a}rbara and V{\'a}squez, Christofer and Saavedra, Gabriel and B{\'o}rquez-Paredes, Danilo and Jara, Nicol{\'a}s and Olivares, Ricardo and Amjad, Saquib and Leiva, Ariel and Mas-Machuca, Carmen},
  title   = {Benchmarking framework for resource allocation algorithms in multicore fiber elastic optical networks},
  journal = {Journal of Optical Communications and Networking},
  volume  = {16},
  number  = {11},
  pages   = {G11--G27},
  year    = {2024},
  doi     = {10.1364/JOCN.534257}
}

@article{bonati2022openrangym,
  title={OpenRAN Gym: AI/ML development, data collection, and testing for O-RAN on PAWR platforms},
  author={Bonati, Leonardo and Polese, Michele and D’Oro, Salvatore and Basagni, Stefano and Melodia, Tommaso},
  journal={Computer Networks},
  volume={220},
  pages={109502},
  year={2023},
  publisher={Elsevier}
}

@article{wu2020matheuristic,
  author  = {Wu, Xinyun and L{\"u}, Zhipeng and Glover, Fred},
  title   = {A matheuristic for a telecommunication network design problem with traffic grooming},
  journal = {Omega},
  volume  = {90},
  pages   = {102004},
  year    = {2020},
  doi     = {10.1016/j.omega.2018.11.012}
}

@article{gendron2018matheuristics,
  author  = {Gendron, Bernard and Hanafi, Sa{\"\i}d and Todosijevi{\'c}, Raca},
  title   = {Matheuristics based on iterative linear programming and slope scaling for multicommodity capacitated fixed charge network design},
  journal = {European Journal of Operational Research},
  volume  = {268},
  number  = {1},
  pages   = {70--81},
  year    = {2018},
  doi     = {10.1016/j.ejor.2018.01.022}
}

@article{dealmeida2024hybrid,
  author  = {de Almeida, Guilherme Barbosa and de S{\'a}, Elisangela Martins and de Souza, S{\'e}rgio Ricardo and Souza, Marcone Jamilson Freitas},
  title   = {A hybrid iterated local search matheuristic for large-scale single source capacitated facility location problems},
  journal = {Journal of Heuristics},
  volume  = {30},
  number  = {3},
  pages   = {145--172},
  year    = {2024},
  doi     = {10.1007/s10732-023-09524-9}
}

@misc{macrae2019hybrid,
  author        = {MacRae, Cameron A.~G. and Ozlen, Melih and Ernst, Andreas T.},
  title         = {A hybrid Benders decomposition and bees algorithm matheuristic approach to transmission expansion planning considering energy storage},
  year          = {2019},
  eprint        = {1903.01236},
  archivePrefix = {arXiv},
  primaryClass  = {math.OC},
  url           = {https://arxiv.org/abs/1903.01236}
}

@article{zheng2025improved,
  author  = {Zheng, Weichang and Ke, Ming and Li, Jinke and Yang, Mingcong and Zheng, Yu and Zhang, Yongbing and Yang, Kun},
  title   = {Improved {ILP} models and heuristics for solving routing and resource allocation problems in optical networks},
  journal = {Journal of Lightwave Technology},
  volume  = {43},
  number  = {7},
  pages   = {3016--3033},
  year    = {2025},
  doi     = {10.1109/JLT.2024.3519778}
}
